\documentclass[twocolumn]{aastex63}

\usepackage{textcomp, gensymb}
\usepackage{xcolor}
\usepackage{wrapfig}
\usepackage{tabularx}
\usepackage{changepage}
\let\tablenum\relax
\usepackage{siunitx}
\usepackage{graphicx,bm,times,amsmath,array,float}
\received{\today}

\shorttitle{VLA CO(1--0)}
\shortauthors{Frias Castillo et al.}

\graphicspath{{./}{figures/}}

\begin{document}

\title{VLA Legacy Survey of Molecular Gas in Massive Star-forming Galaxies at High Redshift}

\correspondingauthor{Marta Frias Castillo}
\email{friascastillom@strw.leidenuniv.nl}

\author[0000-0002-9278-7028]{Marta Frias Castillo}
\affiliation{Leiden Observatory, Leiden University, P.O. Box 9513, 2300 RA Leiden, The Netherlands}

\author[0000-0001-6586-8845]{Jacqueline Hodge}
\affiliation{Leiden Observatory, Leiden University, P.O. Box 9513, 2300 RA Leiden, The Netherlands}

\author[0000-0002-1383-0746]{Matus Rybak}
\affiliation{Faculty of Electrical Engineering, Mathematics and Computer Science,
Delft University of Technology, Mekelweg 4, 2628 CD Delft, the Netherlands}
\affiliation{Leiden Observatory, Leiden University, P.O. Box 9513, 2300 RA Leiden, The Netherlands}

\author[0000-0001-5434-5942]{Paul van der Werf}
\affiliation{Leiden Observatory, Leiden University, P.O. Box 9513, 2300 RA Leiden, The Netherlands}

\author[0000-0003-3037-257X]{Ian Smail}
\affiliation{Centre for Extragalactic Astronomy, Department of Physics, Durham University, South Road, Durham DH1 3LE, UK}

\author[0000-0002-3272-7568]{Jack E. Birkin}
\affiliation{Centre for Extragalactic Astronomy, Department of Physics, Durham University, South Road, Durham DH1 3LE, UK}

\author{Chian-Chou Chen}
\affiliation{Academia Sinica Institute of Astronomy and Astrophysics (ASIAA), No.1, Section 4, Roosevelt Road, Taipei 10617, Taiwan}

\author{Scott C. Chapman}
\affiliation{Department of Physics and Atmospheric Science, Dalhousie University, Halifax, Halifax, NS B3H 3J5, Canada}

\author{Ryley Hill}
\affiliation{Department of Physics and Astronomy, University of British Columbia, 6225 Agricultural Road, Vancouver V6T 1Z1, Canada}

\author[0000-0003-3021-8564]{Claudia del P. Lagos}
\affiliation{International Centre for Radio Astronomy Research (ICRAR), M468, University of Western Australia, 35 Stirling Hwy, Crawley, WA 6009, Australia}
\affiliation{ARC Centre of Excellence for All Sky Astrophysics in 3 Dimensions (ASTRO 3D)}
\affiliation{Cosmic Dawn Center (DAWN), R\aa dmandsgade 62, DK-2200 København, Denmark}

\author[0000-0002-5247-6639]{Cheng-Lin Liao}
\affiliation{Academia Sinica Institute of Astronomy and Astrophysics (ASIAA), No.1, Section 4, Roosevelt Road, Taipei 10617, Taiwan}
\affiliation{Graduate Institute of Astrophysics, National Taiwan University, Taipei 10617, Taiwan}

\author[0000-0001-9759-4797]{Elisabete da Cunha}
\affiliation{International Centre for Radio Astronomy Research, University of Western Australia, 35 Stirling Hwy, Crawley, WA 6009, Australia}
\affiliation{ARC Centre of Excellence for All Sky Astrophysics in 3 Dimensions (ASTRO 3D)}

\author{Gabriela Calistro Rivera}
\affiliation{European Southern Observatory (ESO), Karl-Schwarzschild-Stra\ss e 2, 85748 Garching bei M\"unchen, Germany }

\author[0000-0003-3921-3313]{Jianhang Chen}
\affiliation{European Southern Observatory (ESO), Karl-Schwarzschild-Stra\ss e 2, 85748 Garching bei M\"unchen, Germany }

\author{E.F. Jim\'enez-Andrade}
\affiliation{Instituto de Radioastronom\'ia y Astrof\'isica, Universidad Nacional Aut\'onoma de M\'exico, Antigua Carretera a P\'atzcuaro \# 8701,\\ Ex-Hda. San Jos\'e de la Huerta, Morelia, Michoac\'an, M\'exico C.P. 58089}
\affiliation{National Radio Astronomy Observatory, 520 Edgemont Road, Charlottesville, VA 22903, USA}

\author[0000-0001-7089-7325]{Eric J.\,Murphy}
\affiliation{National Radio Astronomy Observatory, 520 Edgemont Road, Charlottesville, VA 22903, USA}

\author{Douglas Scott}
\affiliation{Department of Physics and Astronomy, University of British Columbia, 6225 Agricultural Road, Vancouver V6T 1Z1, Canada}

\author{A.M. Swinbank}
\affiliation{Centre for Extragalactic Astronomy, Department of Physics, Durham University, South Road, Durham DH1 3LE, UK}

\author[0000-0003-4793-7880]{Fabian Walter}
\affiliation{Max Planck Institute for Astronomy, K\"onigstuhl 17, 69117,  Heidelberg, Germany}
\affiliation{National Radio Astronomy Observatory, Pete V. Domenici Array Science Center, P.O. Box O, Socorro, NM 87801, USA}

\author{R.J. Ivison}
\affiliation{European Southern Observatory (ESO), Karl-Schwarzschild-Stra\ss e 2, 85748 Garching bei M\"unchen, Germany }
\affiliation{Institute for Astronomy, University of Edinburgh, Royal Observatory, Edinburgh EH9 3HJ, UK}

\author[0000-0001-7147-3575]{Helmut Dannerbauer}
\affiliation{Instituto Astrof\'isica de Canarias (IAC), E-38205 La Laguna, Tenerife, Spain}
\affiliation{Dpto. Astrof\'isica, Universidad de la Laguna, E-38206 La Laguna, Tenerife, Spain}

\begin{abstract}
We present initial results of an ongoing survey with the \textit{Karl G. Jansky} Very Large Array  targeting the CO($J$ = 1--0) transition in a sample of 30 submillimeter-selected, dusty star-forming galaxies at $z =$ 2--5 with existing mid--$J$ CO detections from ALMA and NOEMA, of which 17 have been fully observed. We detect CO(1--0) emission in 11 targets, along with three tentative ($\sim$1.5--2$\sigma$) detections; three galaxies are undetected. Our results yield total molecular gas masses of 6--23$\times$10$^{10}$ ($\alpha_\mathrm{CO}$/1)  M$_\odot$, with gas mass fractions, $f_\mathrm{gas}$=$M_\mathrm{mol}$/($M_*$+$M_\mathrm{mol}$), of 0.1--0.8 and a median depletion time of (140$\pm$70) Myr. We find median CO excitation ratios of $r_{31}$ = 0.75$\pm$0.39 and $r_{41}$ = 0.63$\pm$0.44, with a significant scatter. We find no significant correlation between the excitation ratio and a number of key parameters such as redshift, CO(1--0) line width or $\Sigma_\mathrm{SFR}$. We only find a tentative positive correlation between $r_{41}$ and the star-forming efficiency, but we are limited by our small sample size. Finally, we compare our results to predictions from the SHARK semi-analytical model, finding a good agreement between the molecular gas masses, depletion times and gas fractions of our sources and their SHARK counterparts. Our results highlight the heterogeneous nature of the most massive star-forming galaxies at high-redshift, and the importance of CO(1--0) observations to robustly constrain their total molecular gas content and ISM properties.
    
\end{abstract}


\section{Introduction}

Tracing the evolution of the molecular gas content in galaxies is necessary for a complete understanding of galaxy formation and evolution, as it provides the direct fuel for star formation (see reviews by \citealt{kennicutt+evans2012,carilli-walter2013,tacconi2020}). The main component of the molecular gas, molecular hydrogen (H$_2$), cannot be excited in its rotation/vibration transitions in the low temperatures of the interstellar medium (ISM) of galaxies due to the large separation between its energy levels ($\sim$500 K). Instead $^{12}$CO, the second most abundant molecule in galaxies after H$_2$, has been traditionally used to trace the kinematics, dynamics and physical conditions of the cool gas in the ISM. The low upper level temperature $T_\mathrm{ex}$ = 5.5 K and critical density of the rotational ground state ($J = 1 - 0$) of $^{12}$CO, hereafter CO(1--0), means that this molecule is easily excited in a variety of galaxy environments, making it a  useful tool for tracing the bulk of the cold molecular gas. The use of CO(1--0) only requires the assumption of a conversion factor, $\alpha_\mathrm{CO}$, to obtain the total cold molecular gas mass from the CO(1--0) luminosity \citep[for a review, see][]{bolatto2013}.
Dust emission \citep[][]{hildebrand1983,scoville2016,liu2019,wang2022} and the optically thin emission lines from neutral atomic carbon  ([\ion{C}{1}], \citealt{weiss2005b,walter2011,valentino2018}) are often used as alternative tracers of the molecular gas alongside CO. 

Thanks to the improved capacities of (sub)-mm interferometers such as the Atacama Large Millimeter/submillimeter Array (ALMA), the Karl G. Jansky Very Large Array (VLA) or the NOrthern Extended Millimiter Array (NOEMA), studies of the redshifted CO emission have become common at high-redshift \citep{carilli-walter2013,hodge+dacunha2020}. Blind CO line surveys such as the VLA CO Luminosity Density at High Redshift \citep[COLDz, ][]{pavesi2018,riechers2019} and the ALMA Spectroscopic Survey in the Hubble Ultra Deep Field \citep[ASPECS, ][]{walter2016,decarli2019}, have begun to unveil the CO excitation, molecular gas content and physical conditions of the ISM in star-forming galaxies (SFGs) at $z \sim$ 1--6, establishing the most reliable evolution of the cosmic molecular gas density to date \citep[][]{decarli2020,riechers2020a-vlaspecs}.

Obtaining direct observations of the CO(1--0) emission line at high-redshift is a challenging task that requires long integration times, mostly due to the limitations of existing instrumentation in the current facilities in use. Therefore, studies commonly rely on the brighter mid- and high-$J$ CO lines as alternative molecular gas tracers \citep[e.g.,][]{bothwell2013,daddi2015,yang2017,boogaard2020,birkin2020}. This requires an assumption on the shape of the CO spectral line energy distribution (SLED) to infer the CO(1--0) luminosity. The CO excitation depends however on physical conditions and heating mechanisms at play in the cold ISM, and conversion factors span a wide range of values over the high-redshift galaxy population \citep{carilli-walter2013,narayanan_krumholz2014,sharon2016,yang2017,harrington2018,boogaard2020,riechers2020a-vlaspecs}.

As noted, detecting CO(1--0) at high-redshift requires several hours per source for even the brightest systems, the cold-gas-rich submillimeter galaxies (SMGs) \citep{casey2014,hodge+dacunha2020}. These dusty, high-infrared-luminosity ($L_\mathrm{IR} >$ 10$^{12}$ L$_\odot$; \citealt{magnelli2012lir_smg}) systems have a peak cosmic volume density around $z \sim$ 2 -- 3 \citep{chapman2005,danielson2017,Dudzeviciute2020}. They are some of the most active star-forming systems in the Universe, with star-formation rates in the range 100 -- 1000 M$_\odot$ yr$^{-1}$ \citep{magnelli2012lir_smg,dacunha2015,Dudzeviciute2020}, fed by large molecular gas reservoirs of 10$^{10}$ -- 10$^{11}$ M$_\odot$ \citep{greve2005,tacconi2008,bothwell2013,birkin2020}, and the brighter systems tend to be located at higher redshifts \citep{AS2COSPEC2022}. Star formation in SMGs is typically distributed in dust structures with diameters of 2 -- 3 kpc \citep{ikarashi2015,simpson2015a,gullberg2019,hodge2019}. Such intense star-forming episodes are thought to be mainly triggered by mergers or interactions with  neighbouring galaxies. To date, CO(1--0) observations have been preferentially obtained towards bright, strongly lensed systems \citep[e.g.,][]{danielson2011,thompson2012,aravena2013,harrington2018}. Deriving the intrinsic CO(1--0) properties of lensed galaxies is however subject to uncertainties arising from lens modeling. 

Our knowledge of the cold molecular gas content in non-lensed systems comes from targeted studies \citep{carilli2010gn20sled,ivison2011,riechers2011,riechers2011b,sharon2016,huynh2017,kaasinen2019,leung2019,friascastillo2022,xiao2022}. While valuable, carrying out a systematic study of a statistically significant sample of CO(1--0) emission of high-redshift galaxies has been hampered by the rather heterogeneous selection criteria applied in the different studies.

We have therefore undertaken a CO(1--0) survey of 30 unlensed, high-redshift ($z$ = 2--5) submillimeter-selected star-forming galaxies with precise redshifts and existing moderate-$J$ CO line detections from ALMA or NOEMA \citep[][Chapman et al., in prep]{birkin2020,AS2COSPEC2022}. This survey more than doubles the existing number of unlensed $z >$ 2 star-forming galaxies detected in CO(1--0). The paper is organised as follows: we review the sample selection and VLA observations in Section \ref{sec:observations}. In Section \ref{sec:Results} we present the method used for extracting the integrated line fluxes. We then analyse the molecular gas mass content and excitation conditions of the sources in Section \ref{sec:analysis} and we present our conclusions in Section \ref{sec:conclusions}. Throughout this paper we assume a standard $\Lambda$CDM cosmology with $H_0$ = 67.8 km s$^{-1}$ Mpc$^{-1}$, $\Omega_\mathrm{M}$ = 0.310 and $\Omega_\Lambda$ = 0.690 \citep[][]{planck2016}.

\begin{table*}
\centering
\caption{Target sample \label{tab:sample}  and details of JVLA observations}
\begin{tabular}{@{}lcccccccc @{}}
 \hline \hline
Target & R.A. J2000 & Dec J2000 & $z^a$ & Date & rms channel$^{-1}$ $^b$ & Beam & Phase calibrator & Flux calibrator \\
& [hh:mm:ss.ss] & [deg:mm:ss.ss] & &  & [$\mu$ Jy beam$^{-1}$] & [maj $\times$ min, PA] & &\\
\hline
AS2COS0008.1 & 10:02:49.2 & +02:32:55.5 & 3.581 & 18-05-2021 & 87 & 4.6$''\times$3.0$''$,38\degree & J1024-0052 & 3C286\\
& & & & 18-05-2021 &  &  &  & \\
AS2COS0009.1 & 10:00:28.7 & +02:32:03.6 & 2.260 & 17-05-2021 & 92 & 2.9$''\times$2.2$''$,27\degree & J1024-0052 & 3C286\\
& & & & 26-05-2021 &  &  &  & \\
AS2COS0013.1 & 10:00:35.3 & +02:43:53.0 & 2.608 & 10-03-2017 & 76 & 3.2$''\times$2.3$''$,$-$39\degree & J1041+0610 & 3C286\\
AS2COS0023.1 & 09:59:42.9 & +02:29:38.2 & 4.341 & 11-05-2021 & 63 & 4.4$''\times$3.2$''$,$-$6\degree & J1024--0052 & 3C286 \\
AS2COS0031.1 & 09:59:23.0 & +02:51:37.5 & 3.643 & 20-04-2021 & 58 & 4.5$''\times$2.8$''$,13\degree & J1024--0052 & 3C286\\
AS2COS0054.1 & 09:58:45.9 & +02:43:29.3 & 3.174 & 04-04-2021 & 86 & 3.8$''\times$2.5$''$,8\degree & J1024--0052 & 3C286\\
AS2UDS010.0 & 02:15:55.9 & $-$4:55:08.6 & 3.169 & 21-05-2021 & 98 & 4.7$''\times$2.6$''$,$-$42\degree & J0215--0222 & 3C48\\
& & & & 30-05-2021 &  &  & &  \\
AS2UDS011.0 & 02:16:30.8 & $-$5:24:03.3 & 4.073 & 17-04-2021 & 84 & 4.9$''\times$3.0$''$,$-$12\degree & J0215--0222 & 3C48 \\
AS2UDS012.0 & 02:18:03.6 & $-$4:55:27.2 & 2.520 & 25-05-2021 & 65 & 3.5$''\times$2.4$''$,$-$37\degree & J0215--0222 & 3C48\\
AS2UDS014.0 & 02:17:44.3 & $-$5:20:08.6 & 3.804 & 04-05-2021 & 81 & 4.7$''\times$3.0$''$,$-$11\degree & J0215--0222 & 3C48\\
AS2UDS026.0 & 02:19:02.1 & $-$5:28:56.9 & 3.296 & 18-05-2021 & 88 & 3.8$''\times$2.6$''$,$-$16\degree & J0215--0222 & 3C48\\
AS2UDS126.0 & 02:15:46.7 & $-$5:18:49.2 & 2.436 & 22-05-2021 & 92 & 4.4$''\times$2.1$''$,$-$45\degree & J0215--0222& 3C48\\
& & & & 29-05-2021 & &  &  &  \\
AEG2 & 14:15:57.5 & +52:07:12.4 & 3.668 & 01-04-2021 & 56 & 3.2$''\times$2.7$''$,29\degree & J1419+5423 & 3C286 \\
AEG3 & 14:15:47.1 & +52:13:48.4 & 4.032 & 02-04-2021 & 41 & 4.2$''\times$2.9$''$,17.\degree & J1419+5423 & 3C286 \\
CDFN1 & 12:35:55.9 & +62:22:39.2 & 3.159 & 29-03-2021 & 90 & 2.9$''\times$2.5$''$,39\degree & J1400+6210 & 3C286\\
CDFN2 & 12:35:51.5 & +62:21:47.4 & 4.422 & 31-03-2021 & 63 & 4.3$''\times$3.5$''$,0\degree & J1400+6210 & 3C286\\
CDFN8 & 12:36:27.2 & +62:06:05.8 & 4.144 & 28-03-2021 & 75 & 3.7$''\times$2.9$''$,4\degree & J1400+6210 & 3C286\\
\hline
\multicolumn{9}{l}{$^a$ Obtained from mid--$J$ CO line detections} \\
\multicolumn{9}{l}{$^b$ For 100 km s$^{-1}$ channels} \\
 \end{tabular}
\end{table*}

\section{Observations and data reduction}
\label{sec:observations}
\subsection{Sample}

\begin{figure}
    \centering
    \hspace{-0.5cm}
    \includegraphics[width=\linewidth]{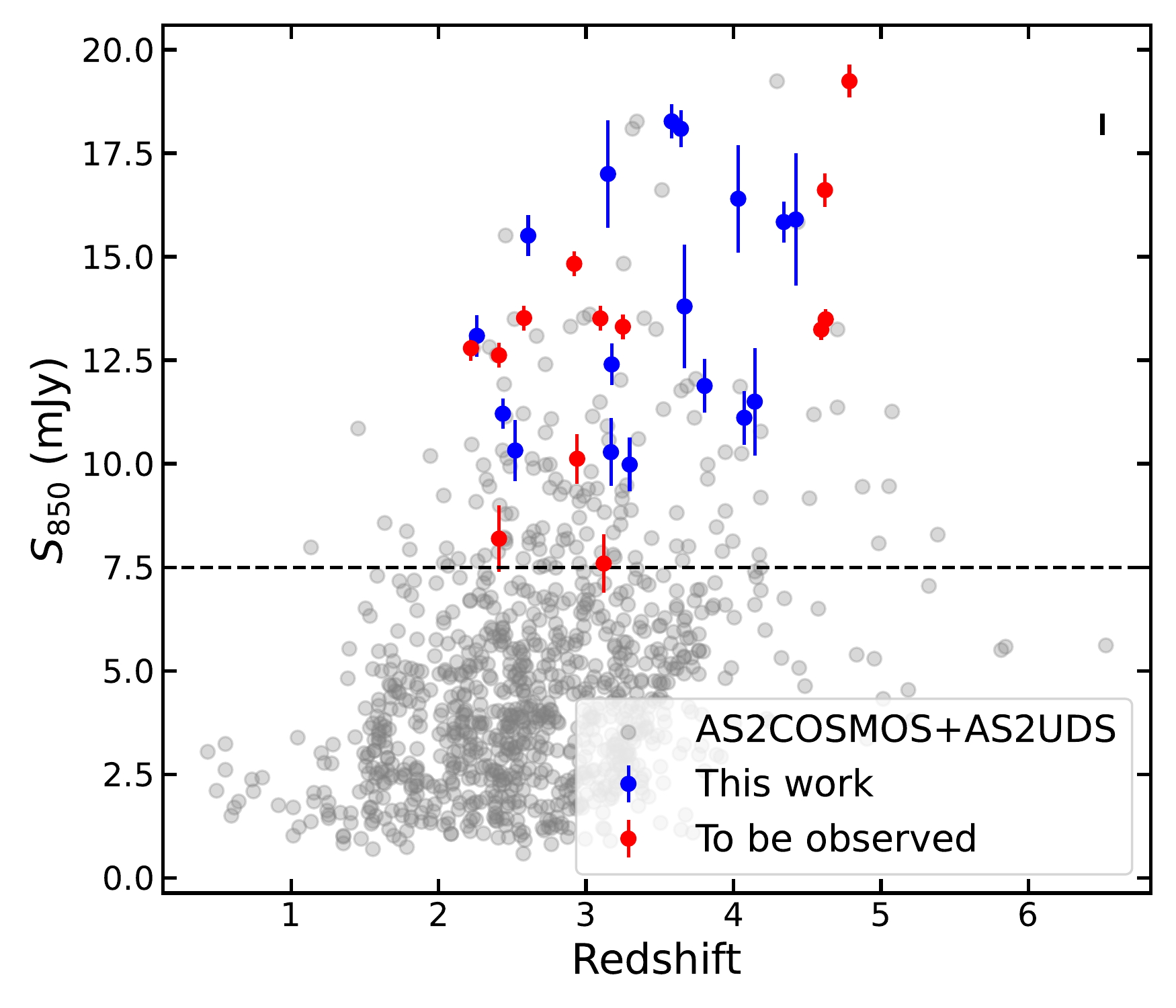}
    \caption{$S_{850}$ flux density versus redshift for the sources presented in this paper (blue circles), with the parent samples from AS2COSMOS \citep{simpson2020} and AS2UDS \citep{stach2019,Dudzeviciute2020} shown as the gray dots. The sources still to be observed in our survey are shown as red dots. Given the time required to detect CO(1--0) in unlensed sources, we have targeted sources with  $S_{850}>$7.5 mJy. The majority of the parent sample only have photometric redshifts, so we further constrained the sample to those targets with spectroscopic redshifts of $z>$1.88 from at least one CO spectral line \citep{birkin2020,AS2COSPEC2022}. The final sample spans a redshift range of $z=$ 2--5 and a dust mass range of 1--10$\times$10$^9$ M$_\odot$}
    \label{sample}
\end{figure}

Our targets are selected from a sample of sources detected within 4 deg$^2$ of SCUBA-2 imaging of the UKIDSS Ultra Deep Survey (UDS), Cosmological Evolution Survey (COSMOS), Chandra Deep Field North (CDFN), and Extended Groth Strip (EGS) fields \citep{geach2017,simpson2019}. The brightest submillimeter sources in these fields were subsequently followed up with ALMA (AS2UDS, \citealt{stach2018}; AS2COSMOS, \citealt{simpson2020}) and SMA \citep[EGS, CDFN,][]{hill2018} continuum imaging and further targeted with blind line scans using ALMA or NOEMA to obtain precise redshifts \citep[][Chapman et al. in prep]{birkin2020,AS2COSPEC2022}. This provided initial constraints on the molecular gas content of a sample of 44 galaxies via mid-$J$ CO transitions. From this parent sample, we selected sources at $z >$ 1.88 to ensure the CO(1--0) line is redshifted into the VLA K- and Ka bands. We excluded sources in the EGS field which only had a single transition detected and thus relied on photometric redshifts. Finally, we selected the 30 brightest sources based on mid-$J$ CO and 850 $\mu$m flux densities ($S_\mathrm{850}> $7.5 mJy, Figure \ref{sample}) with robust spectroscopic redshifts to follow up in CO(1--0) with the VLA. Because 850 $\mu$m flux density selects primarily on dust mass \citep[e.g.,][]{hayward2011,liang2018}, the brightness of these sources effectively provides a cold dust-mass-selected sample of high-redshift galaxies \citep{Dudzeviciute2020}. We currently do not have constrains on the presence of AGN within our sample. However, studies have found that around 15\% of field SMGs host AGN \citep{wang2013}, so we would expect 4$-$5 of our targets to have AGN activity
. The final sample spans a range of redshift ($z$ = 2 -- 5) and dust mass ($M_\mathrm{d}$ =1-10$\times$10$^9$ M$_\odot$). Here we present the initial results of the survey, analysing the 17 out of 30 targets that have been thus far fully observed.

\subsection{Observations}

We observed the CO(1--0) emission (rest-frame frequency: $\nu_\mathrm{rest}$ = 115.2712 GHz) in 17 galaxies from our sample at $z$ = 2.26 -- 4.42 (VLA program ID:21A-254; PI: Hodge). The observations were carried out between 2021 March 28th and 2021 June 1st under good weather conditions in D array configuration, in either one 4-hour or two 2-hour scheduling blocks per source. We used the K- or Ka-band receivers in combination with the WIDAR correlator configured to 8-bit sampling mode to observe a contiguous bandwidth of 2 GHz (dual polarization) at 2MHz spectral resolution. The largest angular scale of detectable emission is 7.9$''$ and 5.3$''$ for K and Ka band, respectively. Nearby quasars J1024-0052, J0215-022, J1419+5423 or J1400+6210 were observed for complex gain and pointing calibration. For bandpass and flux calibration, one of the quasars 3C286 and 3C48 was observed once per scheduling block (see Table \ref{tab:sample} for details of individual sources). One of the sources, AS2COS0013.1, already had suitable archival VLA CO(1--0) observations carried out on 2017 March 10th in D-configuration (VLA program ID:17A-251; PI: Walter). The data were downloaded from the VLA archive and, after visually inspecting the data, some additional antennas were flagged.

The data were manually processed using \texttt{CASA} 6.4.3 \citep{casa}. Time ranges with poor visibilities were manually flagged. The calibrated visibilities were imaged using the \texttt{tclean} algorithm in \texttt{CASA}. We adopt natural weighting to maximise the signal-to-noise ratio (SNR) of the detections, which resulted in final beam sizes ranging from 2.9$''$ to 4.7$''$ at Full Width Half Maximum (FWHM). The resulting data cubes reach a noise level of 41--98 $\mu$Jy beam$^{-1}$ for channels of 100 km s$^{-1}$ width (see Table \ref{tab:sample}). The cubes are not continuum-subtracted, but we do not detect any continuum emission down to a 2$\sigma$ sensitivity threshold.

\section{Results} \label{sec:Results}
\subsection{Line Detections}

\begin{figure*}
    \vspace{-1.5cm}
    \hspace{-1cm}
    \includegraphics[scale=0.7]{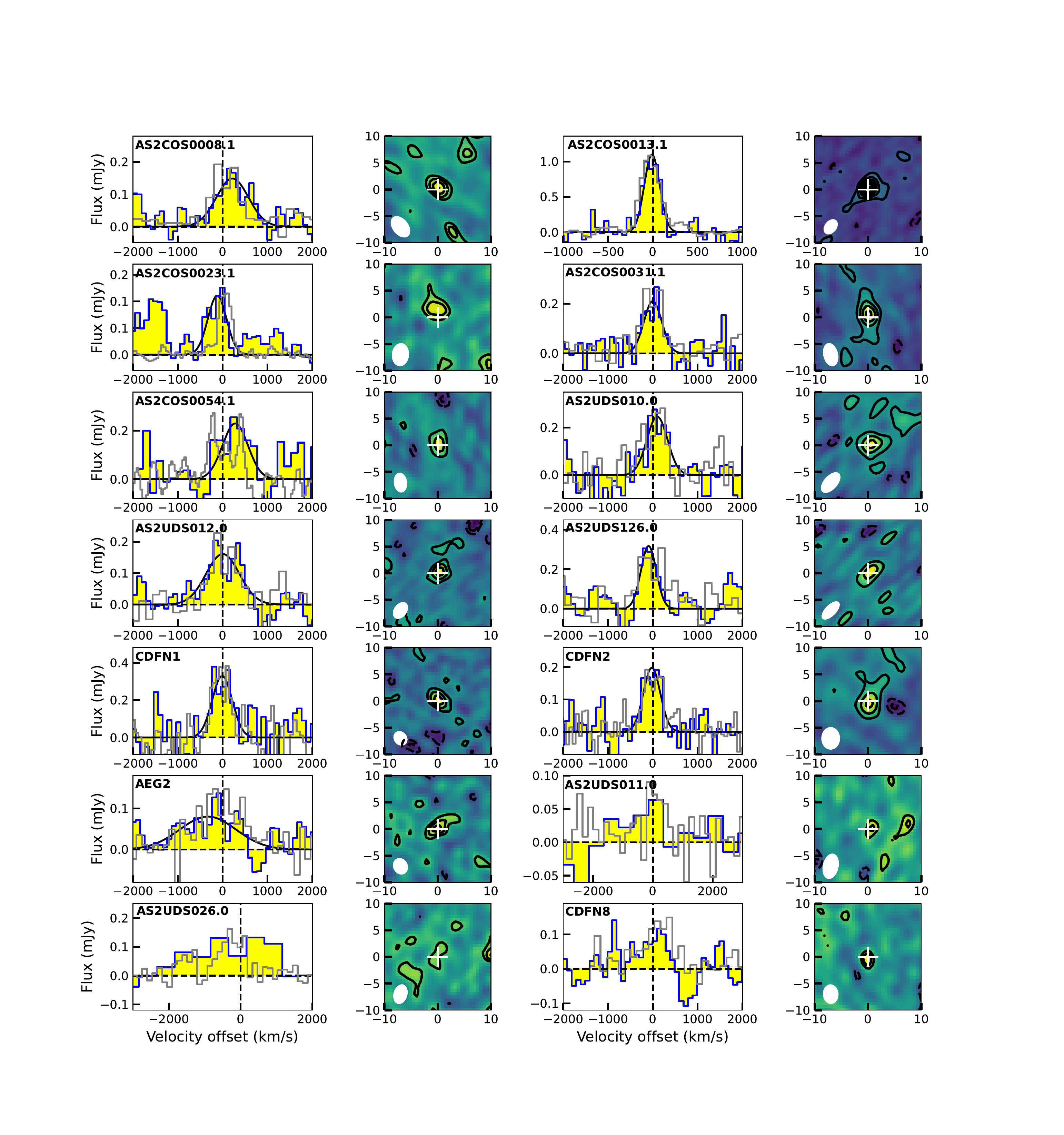}
    \caption{CO(1--0) line emission for the detections and three tentative detections in our sample of SMGs. The spectra (blue line and yellow fill, left panels) are extracted within a 2.5$''$ radius aperture to maximise the SNR. The spectra were fit with a single Gaussian model allowing for a varying line width, shown by the black curve. The 0th-moment maps (right panels) were collapsed over a velocity range equal to the FWHM of the respective mid--$J$ CO emission line and show a 20$''\times$20$''$ field of view. The systemic velocity is based on the redshift derived from the mid--$J$ CO lines, and the gray histograms show the mid-$J$ CO emission line, scaled down in flux density. The white cross indicates the peak of the mid--$J$ CO line emission. Contours start at 2$\sigma$ and increase in steps of 2$\sigma$, except for AS2COS0023.1, CDFN8 and AEG2, where they increase in steps of 1$\sigma$, and AS2COS0013.1, with steps of 3$\sigma$. The white ellipse shows the FWHM of the beam for each source.}
    \label{detections}
\end{figure*}

\begin{figure}
    \centering
    \includegraphics[width=\linewidth]{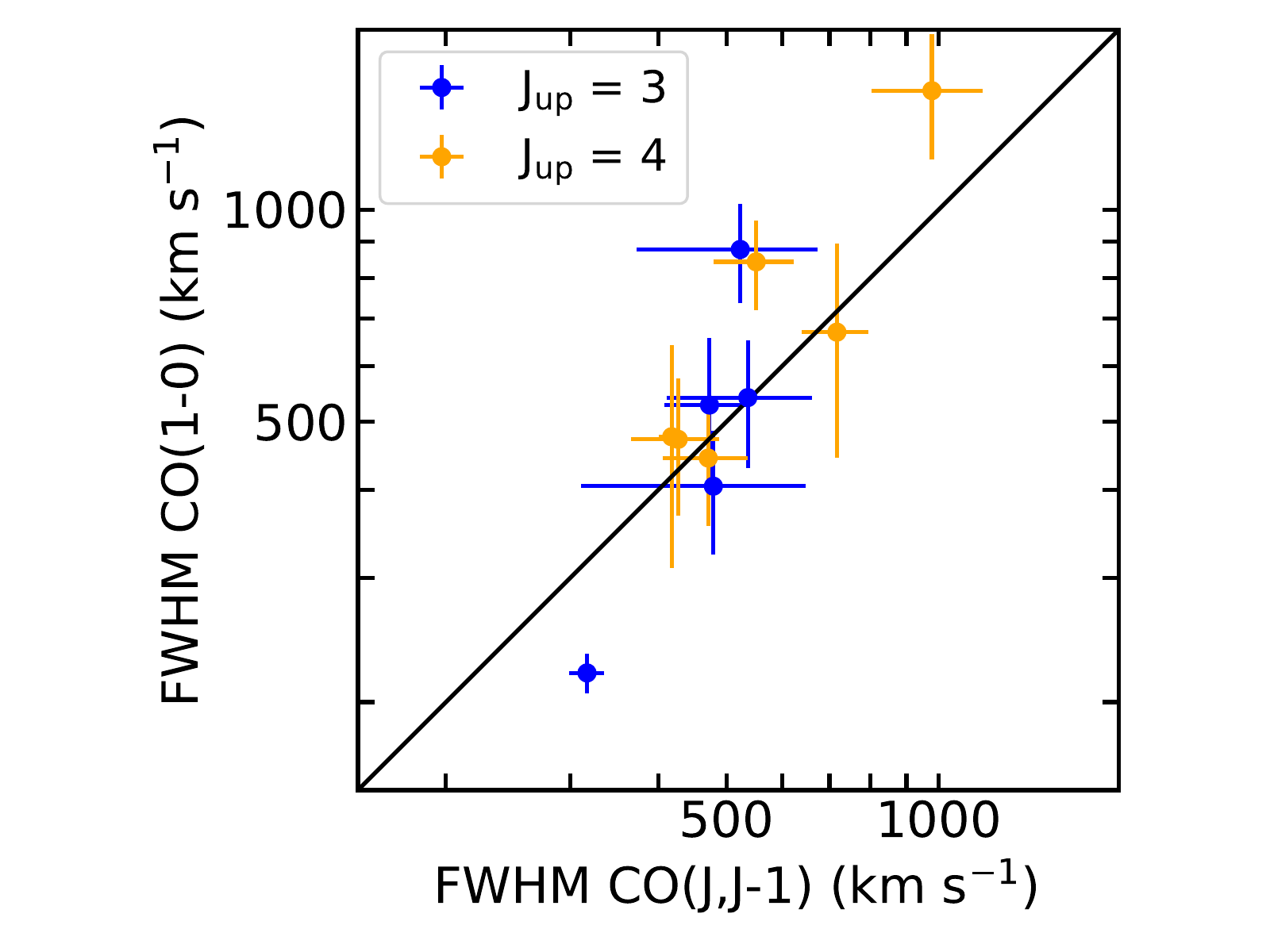}
    \caption{Comparison of the CO(1--0) and CO(3--2) or CO(4--3) line widths for our sample. The solid line shows the one-to-one relation. The line widths agree within 2$\sigma$ for most of our sources, which suggests that, on average, there is not a significant amount of additional cool, diffuse gas being missed by the mid-$J$ transitions. }
    \label{fwhm}
\end{figure}

To search for CO(1--0) emission, we initially create intensity-averaged 0th-moment maps by collapsing the cleaned data cubes over a velocity range corresponding to the FWHM of the respective mid-$J$ CO line detection for each source, using the task \texttt{immoments} in CASA. We detect CO(1--0) towards 11 out of 17 targets at or above the 2$\sigma$ significance level and tentatively detect another three at $\sim$1.5$\sigma$ (Figures \ref{detections}). We do not detect CO(1--0) emission towards AS2COS0009.1, AS2UDS014.0 or AEG3. Using the 0th-moment maps, we check for any spatial offsets between the $J$=1--0 and mid-$J$ CO line emission for the detected sources. This is required for five sources, which show an offset of 0.6$''$, not significant given our resolution (2.9$''$--4.7$''$). Nevertheless, we extract the spectra from the CO(1--0) position for these sources.

We extract the spectra in an aperture of 5$''$ diameter to maximise the SNR and fit them using a single Gaussian model. Figure \ref{fwhm} shows the CO(1--0) line widths against the mid-$J$ CO line widths. Previous studies have suggested that some high-redshift SMGs have line widths in CO(1--0) that are larger than those of the higher--$J$ CO transitions \citep{ivison2010,riechers2011}. Together with the fact that radiative transfer models underpredict the observed low-$J$ CO emission, this suggests the presence of extended, low-excitation gas reservoirs in some SMGs (although not all, e.g. \citealt{hodge2012,friascastillo2022}). We find that the widths of the lines agree with their mid--$J$ counterparts within 2$\sigma$, with a median ratio FWHM$_\mathrm{CO(1-0)}$/FWHM$_\mathrm{CO(J_\mathrm{up},J_\mathrm{up}-1)} =$ 1.1$\pm$0.1. Nevertheless, we have to caution that it is possible that there is fainter emission in the wings of the CO(1--0) lines that is currently not detected due to our sensitivity limitations. Given the sensitivity per channel of our data, and to avoid biasing our analysis, we choose to use the mid-$J$ CO line widths to collapse the data cubes and obtain line fluxes. The integrated line fluxes are consistent within the uncertainties, regardless of whether we use the CO(1--0) or mid-$J$ CO line widths.

To remove any bias due to line structure, we derive line fluxes using the intensity-weighted moments collapsed over a velocity range twice the corresponding mid-$J$ CO line width for each source \citep{bothwell2013,birkin2020}:

\begin{equation}
    M_0 = I_\mathrm{CO} = \int I_\nu d\nu.
\end{equation}

We perform a curve-of-growth analysis on the 0th-moment maps to determine the optimal aperture to extract the line fluxes. In order to increase the SNR, we use the 0th-moment maps collapsed over the velocity range of one full line width ($\pm$0.5$\times$FWHM) of the respective mid-$J$ CO line for each source. We then extract flux densities from a set of circular apertures of increasing diameter, from 1.5$''$ to 40$''$, and determine the point at which the flux converges. In Figure \ref{cog} in the Appendix we show the curves of growth for all the sources as well as for their respective complex gain calibrators. Some of the fainter sources appear not to converge. This is due to the large-scale noise structures, more prominent due to the low SNR of the detections, which has also been observed in other data \citep{novak2020,AS2COSPEC2022}. We note that many of the brighter sources appear to be resolved compared to the phase calibrators, and show extended emission on roughly 6$''$-radius scales. Given the modest SNRs of the detections, we choose to extract the flux from an aperture 2.5$''$ in radius for the sources with SNR$<$3 (integrated over the FWHM of the CO(1--0) line), and then correct to the total flux using a factor of 1.8, as derived from the median curve of growth \citep[following][]{AS2COSPEC2022}. For sources with integrated SNR$>$3, we extract the flux directly from an aperture 6$''$ in radius. The integrated fluxes, line widths and line luminosities are summarised in Table \ref{tab:results}. The 0th-moment maps and spectra of the detection and tentative detections are shown in Figure \ref{detections}. The non-detections are shown in the Appendix.

We note that 6$''$ correspond to radii of $\sim$50 kpc at the median redshift of the sample. Studies of CO(1--0) on a handful of targeted sources have revealed the presence of extended cold gas reservoirs with sizes on the order of a few tens of kpc \citep[][]{ivison2011,emonts2016,dannerbauer2017,frayer2018}. We currently lack the sensitivity and resolution to determine whether all the emission belongs to the same galaxy or if, for example, there could be companions that are contributing to the observed flux. Higher--resolution follow--up is necessary to derive robust sizes and establish the true extent of the gas reservoirs. 
 
Finally, we also note that detectability of molecular gas emission is systematically affected by the increase in the temperature of the Cosmic Microwave Background (CMB) with redshift ($T_\mathrm{CMB}$=$T_0$(1+$z$)). As \cite{dacunha2013} showed, these effects become non-negligible when the CMB temperature becomes close to the gas temperature, $T_\mathrm{kin}$. However, \cite{Jarugula2021} and \cite{harrington2021} have recently reported high kinetic temperatures ($T_\mathrm{dust}$ $\sim$ 45 K, $T_\mathrm{kin}$/$T_\mathrm{dust}$ $\approx$ 2.5) for strongly-lensed dusty star-forming galaxies at high redshift, which might suggest that the CMB has a relatively minor effect on the suppression of the observed CO(1--0) luminosity (less than 15\%). Further, we might expect to see a trend in excitation ratio with redshift if the CO(1--0) line was being heavily affected by CMB suppression. However, as we see in Section \ref{sec:analysis}, there is no such trend. Detailed modelling of the CO SLED of each source would be necessary to derive accurate $T_\mathrm{kin}$ and therefore correct both the observed CO(1--0) and mid--$J$ CO line fluxes.

\subsection{SED Fitting}

In order to consistently derive key parameters such as SFRs and stellar masses for all our sources, we fit their SEDs with the high-redshift version of \textsc{Magphys} \citep{dacunha2015, highz_magphys}, fixing the redshift as that corresponding to the mid--$J$ CO transition, as they are higher SNR than our CO(1--0) data. For details of the photometry used, we refer the reader to \cite{simpson2020} for sources in AS2COSMOS and \cite{Dudzeviciute2020} for AS2UDS.

\begin{figure*}[t]
    \centering
    \includegraphics[scale=0.54]{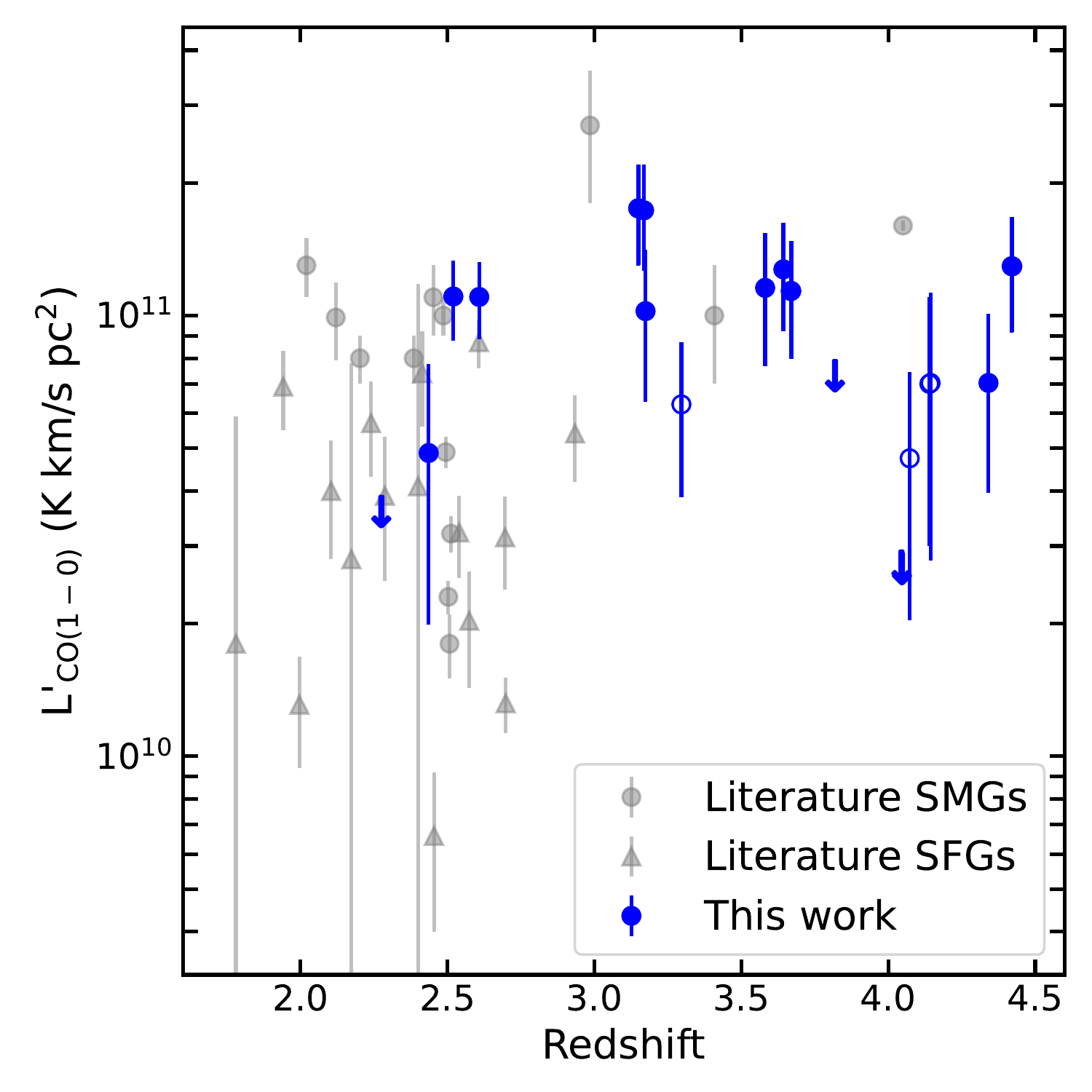}
    \includegraphics[scale=0.55]{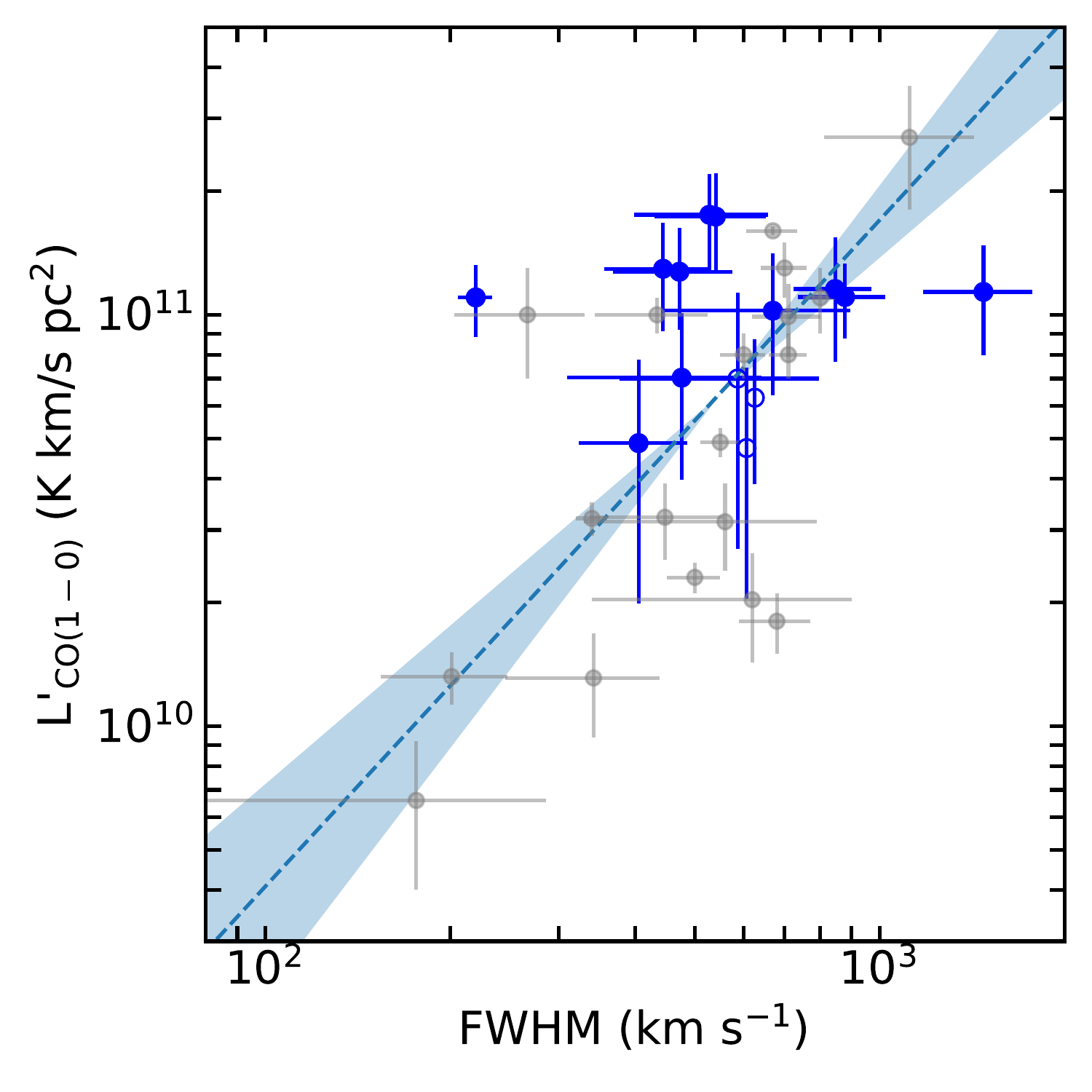}
    \caption{Left: Integrated CO(1--0) line luminosities for our sample as a function of redshift. Open symbols indicate tentative (2$\sigma$) detections and upper limits are marked as downward arrows. We show all non-lensed SMGs with CO(1--0) detections from the literature for comparison, as well as $z$=2 SFGs from \cite{kaasinen2019} and \cite{boogaard2020}. We find no evidence of evolution of $L'_\mathrm{CO(1-0)}$ with redshift. Right:  Integrated CO(1--0) line luminosities as a function of CO(1--0) line width. We fit all the data (including literature values) with the model log$_{10} L'_{CO}$ = $a$ log$_{10}$(FWHM/FWHM$_\mathrm{med}$) + $b$, where FWHM$_\mathrm{med}$ is the median FWHM of all the sources being considered for the fit, 550 km s$^{-1}$. The fit yields a slope $a$ = 1.6$\pm$0.3 and $b$ = 10.8$\pm$0.1.}
    \label{lprimes}
\end{figure*}

For the five sources in the EGS and GOODS-North fields, AEG2, AEG3, CDFN1, CDFN2 and CDFN8, we compiled the available photometry. AEG2 has a counterpart in the DEEP2 Galaxy Redshift Survey photometric catalogs, and we use the published CFHT BRI measurements. For CDFN8, we obtain Ks and IRAC Bands 1 and 2 photometry from \cite{wang2010} and \cite{ashby2015}. At long wavelengths, we use data from the GOODS-Herschel program of \cite{elbaz2011} and the HerMES program of \cite{oliver2012} to measure SPIRE 250$\mu$m, 350$\mu$m, and 500$\mu$m fluxes for sources in the CDFN and EGS fields, respectively. Finally, we use the SMA 870$\mu$m flux density measurements from Chapman et al., (in prep).

CDFN1 and CDFN2 are close to bright foreground sources, and we could not deblend the optical and infrared photometry to obtain stellar masses. Instead, we calculate the total infrared luminosities (8--1000 $\mu$m) by fitting a modified blackbody (MBB) model to the Herschel and ALMA data, and convert to SFR following \cite{kennicutt1998} (correcting for a Chabrier IMF). To check that the SFR values obtained through FIR SED modelling were consistent with those derived from MAGPHYS, we re-fit the FIR SED of the AS2COSMOS and AS2UDS sources and calculated their SFRs from the inferred total infrared luminosities. The median ratio between the MAGPHYS- and FIR-derived SFRs is 1.1 ± 0.2. We therefore consider the MAGPHYS and FIR-derived SFRs to be consistent. We adopt the median stellar mass of our sample for these two sources. Finally, AEG3 is also blended with a foreground source, which is further contaminating IRAC and Herschel fluxes, so we adopt both the median stellar mass and SFR from our sample for this source. Table \ref{tab:lit} presents the final values used in this paper.

\section{Analysis} \label{sec:analysis}

In Section~3, we have described our CO(1--0) observations and the homogenisation of the spectral energy distribution modelling. We now turn towards analysis of this dataset. In Sections~4.1 and 4.2, we discuss the CO(1--0) luminosities and line widths in our sample, infer the molecular gas mass fractions and depletion timescales. We then supplement our sample with literature data where available to study the \textit{general} SMG population. Namely, we assess the reliability of dust-based mass estimates (Section 4.3) and the dependence of the CO excitation of the galaxy properties (Section 4.4). Finally, we compare our sample to the predictions from the SHARK semianalytic models.

\begin{table*}
\caption{Summary of line observations. The columns give source name, 870 $\mu$m flux density, CO(1--0) integrated flux, CO(1--0) FWHM (from a single Gaussian fit to the spectra), peak SNR, CO(1--0) line luminosity, star formation rate, stellar mass, gas fraction and depletion time. Non-detections are reported as 2$\sigma$ upper limits. \label{tab:results} }
\begin{center}
\begin{adjustwidth}{-2.5cm}{}
 \begin{tabular}{@{}lccccccccc@{}}
 \hline \hline
Target & $^aS_{870}$ & $I_\mathrm{CO(1-0)}$ & FWHM$_\mathrm{CO(1-0)}$ & SNR$_\mathrm{peak}$ & $L'_\mathrm{CO(1-0)}$ ($\times$10$^{10}$) & SFR & $M_*$ & $f_\mathrm{gas}$ & $t_\mathrm{dep}$ \\
 & [mJy] & [Jy km s$^{-1}$] & [km s$^{-1}$] & & [K km s$^{-1}$ pc$^{-2}$] & [M$_\odot$ yr$^{-1}$] & [M$_\odot$ ($\times$10$^{10}$)] & & [Myr] \\
\hline
AS2COS0008.1 & 18.3$\pm$0.4 & 0.21 $\pm$ 0.07 & 850$\pm$120 & 2.6 & 11.6 $\pm$ 3.9 & 1400$^{+100}_{-140}$ & 51$^{+12}_{-7}$ & 0.23$\pm$0.07 & 110$\pm$40 \\
AS2COS0009.1 & 13.1$\pm$0.3 & $<$0.14 &  -- & 0.2 & $<$3.6  & 130$^{+3}_{-0}$ &  9$^{+0}_{-1}$ & $<$0.4 & $<$380\\
AS2COS0013.1 & 15.5$\pm$0.4 & 0.34 $\pm$ 0.07 & 220 $\pm$ 14 & 8.8 & 11.0 $\pm$ 2.2 & 920$^{+120}_{-130}$ & 28$^{+8}_{-7}$ & 0.35$\pm$0.08 & 160$\pm$40\\
AS2COS0023.1 & 15.8$\pm$0.3 & 0.09 $\pm$ 0.04 & 480 $\pm$ 170 & 2.1 & 7.0 $\pm$ 3.1 & 590$^{+21}_{-13}$ & 6$^{+0.1}_{-0.1}$ & 0.61$\pm$0.10 & 160$\pm$70\\
AS2COS0031.1 & 18.1$\pm$0.4 & 0.22 $\pm$ 0.06 & 470 $\pm$ 100 & 3.6 & 12.7 $\pm$ 3.5 & 450$^{+100}_{-140}$ & 5$^{+0}_{-0.1}$ & 0.79$\pm$0.05 & 390$\pm$160\\
AS2COS0054.1 & 12.4$\pm$0.2 & 0.23 $\pm$ 0.09 & 670 $\pm$ 220 & 2.7 & 10.2 $\pm$ 3.9  & 1000$^{+120}_{-130}$ & 9$^{+4}_{-2}$ & 0.60$\pm$0.13 &  140$\pm$50\\
AS2UDS010.0 & 10.3$\pm$0.8 & 0.38 $\pm$ 0.10 & 540 $\pm$ 110 & 3.4 & 17.3 $\pm$ 4.7 & 570$^{+90}_{-120}$ & 35$^{+18}_{-14}$ & 0.40$\pm$0.14 & 410$\pm$140 \\
AS2UDS011.0 & 11.1$\pm$0.7 & 0.07 $\pm$0.04 & -- & 1.3 & 4.7 $\pm$ 2.7 & 959$^{+170}_{-190}$ &  32$^{+16}_{-13}$ & 0.17$\pm$0.36 & 67$\pm$40\\
AS2UDS012.0 & 10.3$\pm$0.7 & 0.36 $\pm$ 0.07 & 880 $\pm$ 140 & 3.7 & 11.1 $\pm$ 2.3  & 400$^{+0}_{-5}$ & 21$^{+0}_{-1}$ & 0.42$\pm$0.05 & 380$\pm$80\\
AS2UDS014.0 & 11.9$\pm$0.6 & $<$0.12 & -- & $-$0.2 & $<$7.3 & 690$^{+160}_{-40}$ & 14$^{+2}_{-2}$ & $<$0.4 & $<$140\\
AS2UDS026.0 & 10.0$\pm$0.6 & 0.13 $\pm$ 0.05 & -- & 1.4 & 6.3 $\pm$ 2.9 & 350$^{+80}_{-100}$&  28$^{+12}_{-9}$ & 0.23$\pm$0.42 & 250$\pm$120 \\
AS2UDS126.0 & 11.2$\pm$0.4 & 0.17 $\pm$ 0.10 & 400 $\pm$ 80 & 2.8 & 4.9 $\pm$ 2.9 & 690$^{+340}_{-260}$ & 69$^{+93}_{-44}$ & 0.09$\pm$0.07 & 100$\pm$80\\
AEG2 & 13.8$\pm$1.5 &0.12 $\pm$ 0.06 & 1500 $\pm$ 300 & 2.6 & 11.4 $\pm$ 3.4 & 580$^{+10}_{-10}$ & 5$^{+1.0}_{-0.1}$ & 0.77$\pm$0.06 & 270$\pm$80 \\
AEG3 & 16.4$\pm$1.3 &$<$0.04 & -- & $-$0.1 & $<$2.7 & $^{b}$690$^{+270}_{-120}$ & $^{b}$25$^{+10}_{-20}$ & $<$0.1 & $<$50 \\
CDFN1 & 17.0$\pm$1.3 &0.39 $\pm$ 0.10 & 530 $\pm$ 130 & 4.1 & 17.5 $\pm$ 4.5 & 1150$^{+30}_{-40}$ & $^{b}$25$^{+10}_{-16}$ & 0.49$\pm$012 & 210$\pm$50\\ 
CDFN2 & 15.9$\pm$1.6 &0.17 $\pm$ 0.05 & 440 $\pm$ 90 & 3.1 & 12.9 $\pm$ 3.8 & 1700$^{+80}_{-50}$ & $^{b}$25$^{+10}_{-16}$ & 0.42$\pm$0.12 & 100$\pm$30\\
CDFN8 & 11.5$\pm$1.3 &0.10 $\pm$ 0.06 & 590 $\pm$ 210 & 1.8 & 7.0 $\pm$ 4.3  & 800$^{+180}_{-120}$ & 62$^{+16}_{-18}$ & 0.13$\pm$0.13 &  120$\pm$80 \\
\hline
Median & 13.1$\pm$1.5 & 0.20$\pm$0.04 & 530$\pm$100 & & 10.2$\pm$2.1 & 690$^{+270}_{-120}$ & 25$^{+10}_{-16}$ & 0.35$\pm$0.21 & 140$\pm$70 \\
  \hline
\multicolumn{9}{l}{$^a$ 870$\mu$m flux density measurements obtained from ALMA for sources in the COSMOS and UDS fields \citep{stach2018,simpson2020}}\\
\multicolumn{9}{l}{and from the SMA for those in the CDFN and EGS fields \citep{hill2018} .}\\ 
 \multicolumn{9}{l}{$^b$ Reported values are the median of the whole sample} \\    
 \end{tabular}
 \end{adjustwidth}{}
  \end{center}
\end{table*}

\subsection{CO Line Luminosities}
We convert CO line intensities into line luminosities following \cite{solomon2005}:
\begin{equation}
    L\mathrm{'{_{CO}}} = 3.25 \times 10^7 \ I\mathrm{{_{CO}}} \ \mathrm{\nu{^{-2}_{obs}}} \ D\mathrm{{^2_{L}}} \ (1+\mathrm{z})^{-3} \ \mathrm{K \ km \  s^{-1} \ pc^2} ,
\end{equation}
\noindent where $I_\mathrm{\text{CO}}$ is the integrated line flux from the 0th-moment map in Jy km s$^{-1}$, $\mathrm{\nu_{\text{obs}}}$ is the observed frequency in GHz and $D\mathrm{_{\text{L}}}$ is the luminosity distance in Mpc. 

We find $L'_\mathrm{CO(1-0)}$ luminosities in the range 7 -- 17.5 $\times$ 10$^{10}$ K km s$^{-1}$ pc$^2$, with a median of (10.2$\pm$2.1) $\times$ 10$^{10}$ K km s$^{-1}$ pc$^2$ for the 14 sources. In Figure \ref{lprimes} (left), we show $L'_\mathrm{CO(1-0)}$ against redshift. Tentative detections are shown as open symbols, and non-detections are marked as 2$\sigma$ upper limits. We see no variation with redshift, even after including all the non-lensed SMGs with CO(1--0) detections from the literature (Table \ref{tab:lit}). This stands in contrast with the positive trend for mid-$J$ CO lines found by \cite{birkin2020}. However, since we are targeting the brightest 870$\mu$m-selected sources from the AS2COSMOS, AS2UDS, CDFN and EGS surveys, we are biased towards the most massive systems at their redshifts.

In Figure \ref{lprimes} (right) we show the line luminosities as a function of the CO(1--0) line widths. The $L'_\mathrm{CO}$-FWHM relation serves as a crude probe of the ratio between gas mass and dynamical mass \citep{harris2012,bothwell2013}. Our sources show no correlation between these two parameters, although we are limited by the narrow range in $L'_\mathrm{CO}$. For comparison, we include SMGs and SFGs from the literature with CO(1--0) detections (Table \ref{tab:lit}) and fit all the data with the model log$_{10} \mathrm{L'_{CO}}$ = $a$ log$_{10}$(FWHM/FWHM$_\mathrm{med}$) + $b$, where FWHM$_\mathrm{med}$ is the median FWHM of all the sources being considered for the fit, 550 km s$^{-1}$. The fit yields a significant slope $a$ =
1.6$\pm$0.3 and $b$ = 10.8$\pm$0.1, consistent with the values found from mid-$J$ CO emission \citep{bothwell2013,birkin2020}. 

\subsection{Gas masses, gas fractions and depletion times}

The CO(1--0) emission line is the most direct probe of total molecular gas mass. We calculate the total cold molecular gas mass from the CO(1--0) line luminosities using:
\begin{equation}
M_\mathrm{mol} = 1.36\alpha_\mathrm{CO}L'_\mathrm{CO}\ \mathrm{M_\odot}, 
\end{equation}
where the 1.36 factor accounts for the Helium abundance and $\alpha_\mathrm{CO}$ is the CO-H$_2$ conversion factor in units of M$_\odot$ (K km s$^{-1}$ pc$^2$)$^{-1}$.
This conversion factor depends on several physical parameters, such as temperature, cloud density and metallicity \citep{narayanan2012,bolatto2013}. It is common to assume an ULIRG-like value of $\alpha_\mathrm{CO} \sim$ 1 for starburst systems at high redshift, such as our SMGs. However, recent studies based on dynamical modelling point towards a value of $\alpha_\mathrm{CO}$ = 1--2 \citep{danielson2011,Rivera2018,birkin2020, friascastillo2022}. Assuming $\alpha_\mathrm{CO}$ = 1, we find a median $M_\mathrm{mol}$ = (1.4 $\pm$ 0.3) $\times$ 10$^{11}$ M$_\odot$. Compared to the average SMG population, which has a median of (9.1$\pm$0.7) $\times$ 10$^{10}$ M$_\odot$ \citep{birkin2020}, our targets are almost a factor of two more gas rich. 

\begin{figure*}
    \centering
    \includegraphics[scale=0.5]{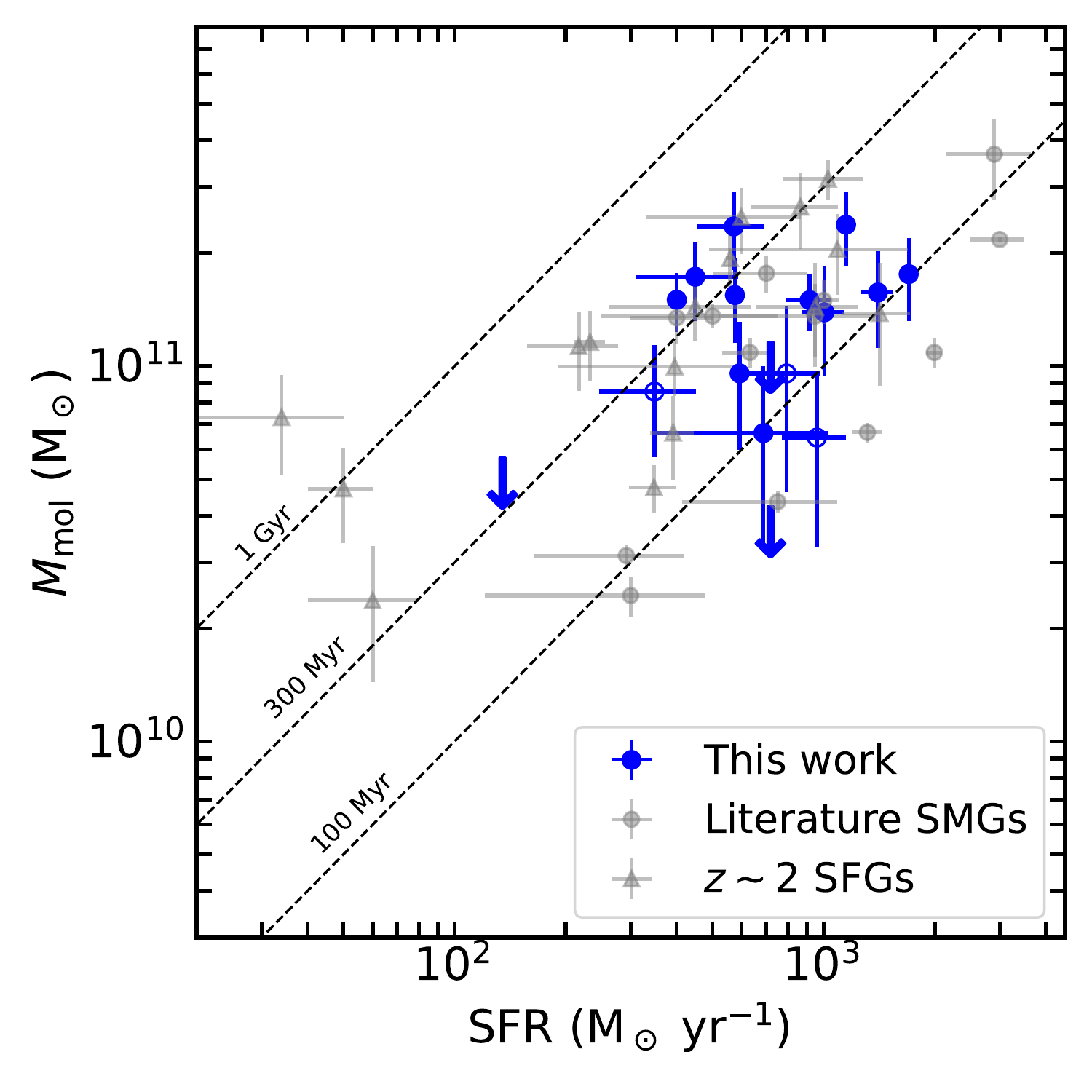}
    \includegraphics[scale=0.5]{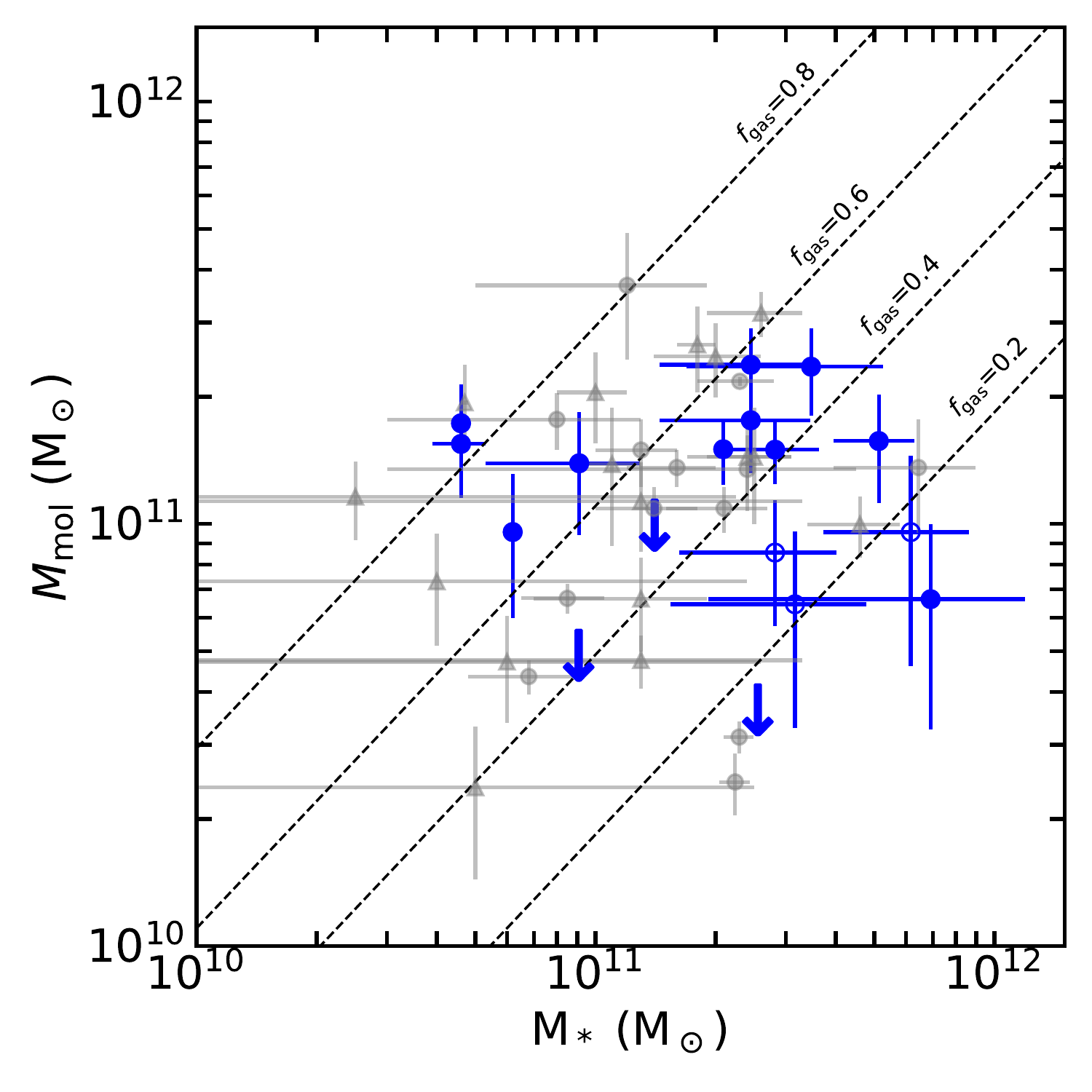}
    \caption{Left: $M_\mathrm{mol}$ vs SFR for our sources, compared to unlensed SMGs with CO(1--0) measurements from the literature, as well as $z\sim$2 SFGs from \cite{kaasinen2019} and \cite{riechers2020a-vlaspecs}. The gas masses from the literature compilation have been adjusted to $\alpha_\mathrm{CO}$=1. The dashed lines show the location of constant gas depletion timescales. Our targets have a median of (140 $\pm$ 70) ($\alpha_\mathrm{CO}$) Myr, in agreement with $t_\mathrm{dep}$ = (210 $\pm$ 40) Myr found from mid--$J$ CO lines for the latest SMG compilations \citep[]{birkin2020}. Right: Same, but in the $M_\mathrm{mol}$-$M_\mathrm{*}$ plane. Dashed lines show the location of constant gas fraction. We find gas fractions in the range 10--80\%, with a median $f_\mathrm{gas}$ of 0.35$\pm$0.21 and a scatter of 0.6 dex.
    }
    \label{mgas}
\end{figure*}

With the gas measurements, we can also explore two key parameters to understand the ISM properties of the galaxies in our sample, the gas fraction and depletion timescale, defined as:
\begin{equation}
  f_\mathrm{gas} = \frac{M_\mathrm{mol}}{M_\mathrm{mol}+M_*}  
\end{equation}
\begin{equation}
t_\mathrm{dep} = \frac{M_\mathrm{mol}}{SFR},
\end{equation}

\noindent which correspond, respectively, to the fraction of baryons available for star formation and  the time that it would take for the systems to use their current gas supply given its current galaxy integrated SFR, in the absence of feedback. 

Figure \ref{mgas} (left) shows the total molecular gas mass as a function of SFR. The dashed lines show the location of constant molecular gas depletion timescale. Our targets have a median of (140 $\pm$ 70) ($\alpha_\mathrm{CO}$) Myr, including tentative detections and upper limits. This is in agreement with $t_\mathrm{dep}$ = (210 $\pm$ 40) Myr found from mid--$J$ CO lines by \cite{birkin2020}, but significantly shorter than the $\sim$1 Gyr that would be expected from scaling relations \citep{tacconi2018} -- although these have been claimed empirically only up to $z\sim$2.5. In Figure \ref{mgas} (right), we show the total molecular gas mass as a function of stellar mass, with the dashed lines showing the location of constant gas fraction. We find gas fractions in the range 10--80\%, with a median $f_\mathrm{gas}$ of 0.35$\pm$0.21. We see a larger amount of scatter in gas fractions than in depletion timescales, driven by the wide range in stellar masses (0.6 dex).

\begin{figure*}[t]
    \centering
    \includegraphics[scale=0.55]{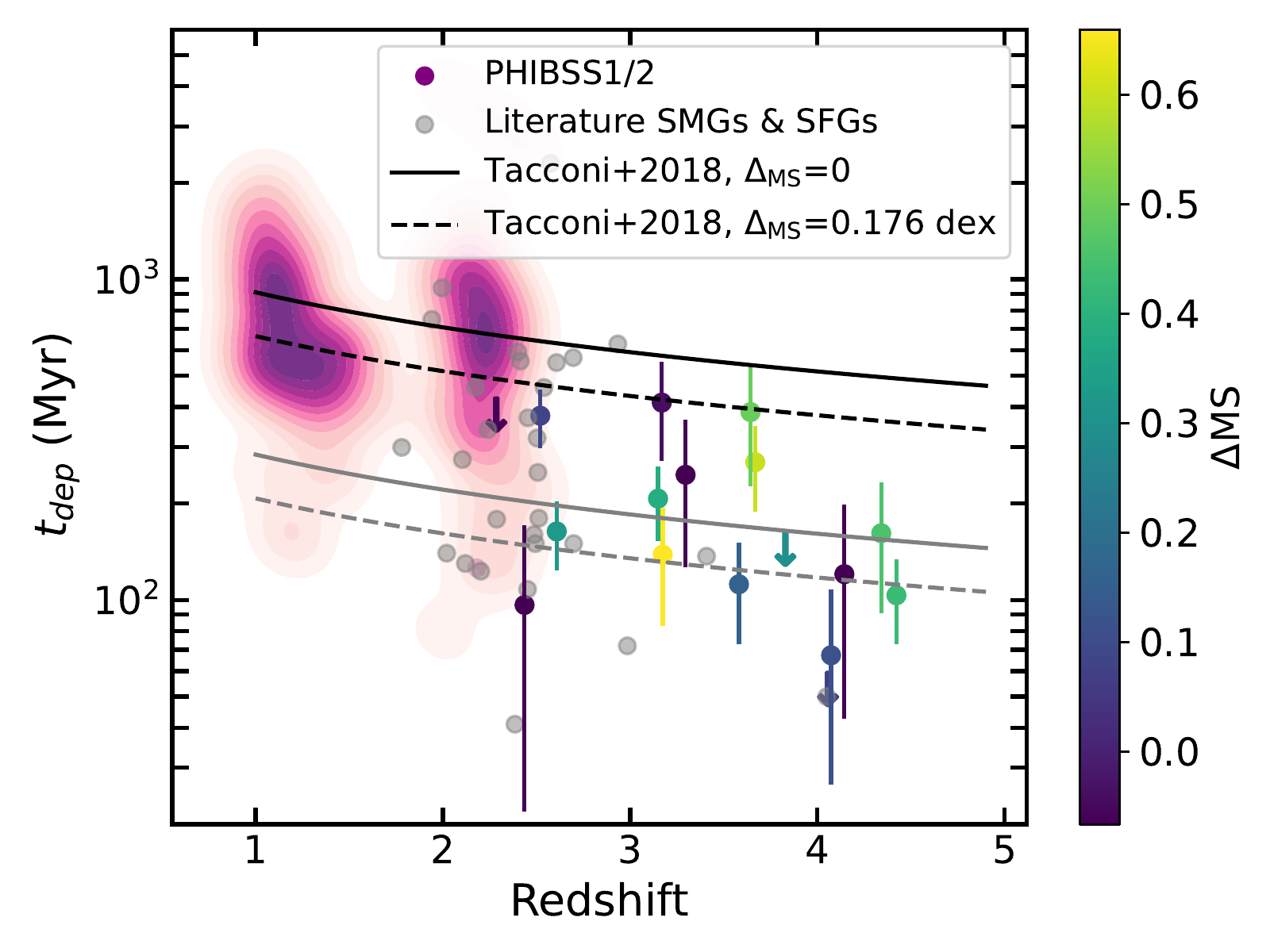}
    \includegraphics[scale=0.55]{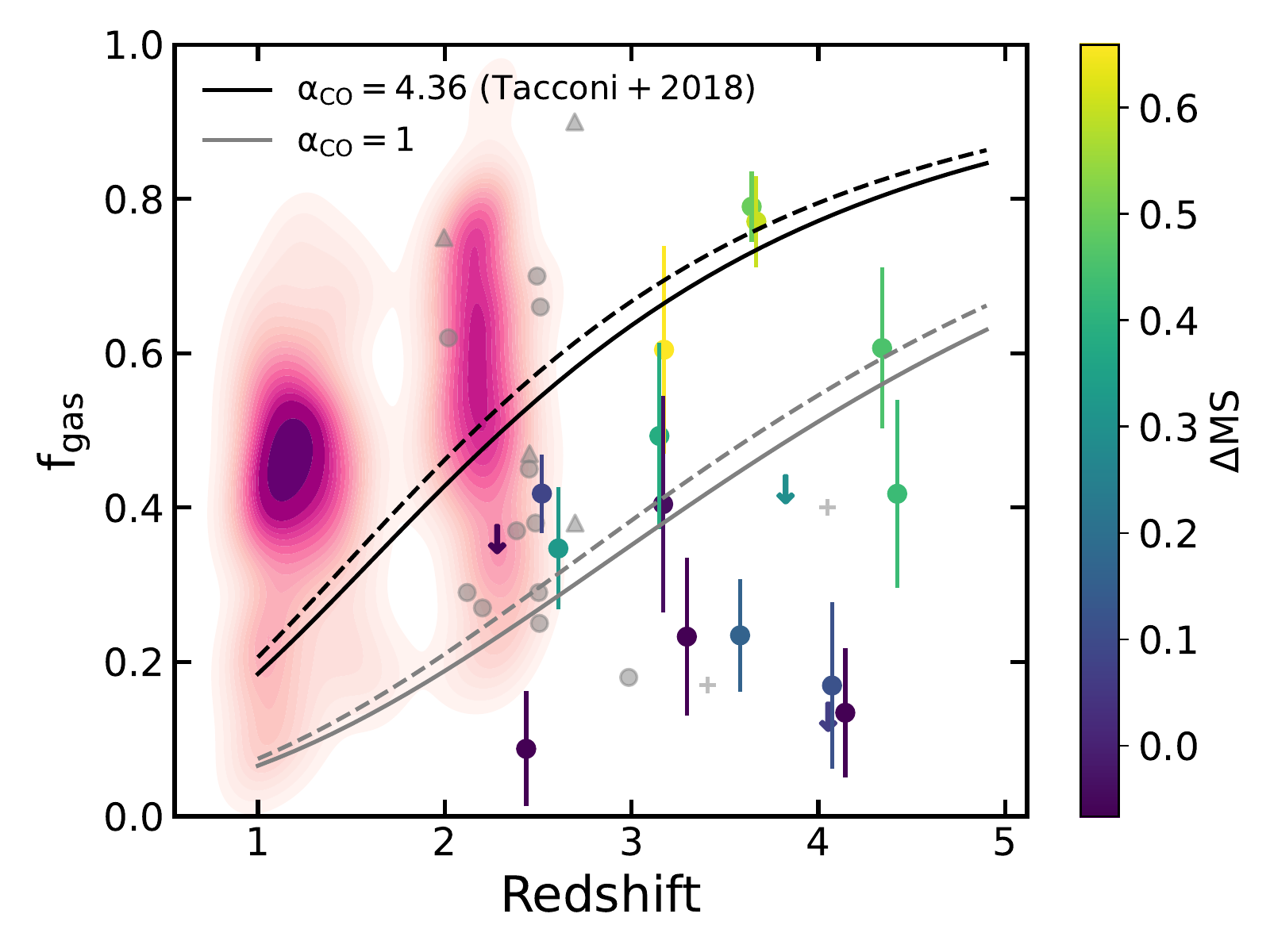}
    \caption{Depletion time (left) and gas fraction (right) as a function of redshift for our SMG sample, colour--coded by their offset from the MS. The solid and dashed lines show the scaling relations from \citet{tacconi2018} for sources on the main sequence and $\times$1.5 above the main sequence, respectively, for a stellar mass of 2.5 $\times$ 10$^{11}$ M$_\odot$, the median stellar mass of our sample. We do not see evidence of evolution with redshift for either parameter. Our gas fractions and depletion times are below the values predicted by the scaling relations, although this disagreement is reduced if we re-scale the relations to $\alpha_\mathrm{CO}$=1 (gray lines). }
    \label{fgas}
\end{figure*}

We also show the evolution of the gas fraction and depletion timescale for our sources in Figure \ref{fgas}, colour-coded by their main sequence (MS) offset, where the MS is defined following \cite{speagle2014}. While there is no trend with MS offset for depletion time, the sources with the largest offset tend to have larger gas fractions, indicating a larger availability of molecular gas to feed the ongoing starburst. Both of these parameters have been shown to follow scaling relations with redshift from previous studies of star-forming galaxies \citep[e.g.,][]{genzel2015,tacconi2018,tacconi2020,liu2019}, shown in Figure \ref{fgas} as solid and dashed lines for sources on and above the main sequence, respectively. Due to the small sample size and large scatter, even after including values from our literature compilation, we refrain from fitting the data. We do not see evolution of $t_\mathrm{dep}$ or $f_\mathrm{gas}$ with redshift for the massive, gas-rich SMGs that comprise our sample. Our gas fractions and depletion times are below the values predicted by the scaling relations for the median stellar mass of our sample. This is partly because \cite{tacconi2018} use a Milky-way value for $\alpha_\mathrm{CO}$ of 4.36 M$_\odot$ (K km s$^{-1}$ pc$^2$)$^{-1}$. Furthermore, the relations were derived from mid--$J$ CO transitions and used excitation correction factors of $r_\mathrm{31}$ and $r_\mathrm{41}$ of 0.42 and 0.31, respectively, to obtain total molecular gas masses. These values are lower than the median for our sources (Section \ref{sec:analysis-excitation}), which results in an overestimation of the total molecular gas mass, and therefore higher gas fractions and depletion times. If we modify the scaling relations to $\alpha_\mathrm{CO} =$ 1 (gray lines in Fig. \ref{fgas}), we find a better agreement with the expected depletion times.

\subsection{Comparison with Dust-based Gas Mass Estimates}

The Rayleigh–Jeans (RJ) tail of dust emission is almost always optically thin, which means it can be used as a tracer of the total dust mass and therefore the molecular gas mass, provided that the dust emissivity per unit mass and the dust-to-gas abundance ratio can be constrained. Under the assumption of a mass-weighted cold dust temperature $T_\mathrm{dust}$ = 25 K (which is claimed to be a representative value for both local star-forming galaxies and high-redshift galaxies) and a dust emissivity index $\beta$ = 1.8, the CO(1--0) luminosity and 850 $\mu$m continuum flux have been shown to correlate for a range of galaxy populations (e.g., \citealt{scoville2016} for local SFGs, ULIRGs and high-redshift SMGs, and \citealt{kaasinen2019} for $z\sim$2 SFGs). 

Using the 870-$\mu$m flux density measurements for our sources, we can estimate the rest-frame $L_\nu$(850 $\mu$m) and therefore $M_\mathrm{mol}$ values following \cite{scoville2016}:

\begin{multline}
 M_\mathrm{mol} = 1.78 S_{\nu_\mathrm{obs}}(1+z)^{-4.8} \times \Big(\frac{\nu_{850}}{\nu_\mathrm{obs}}\Big)^{3.8} \\ D_\mathrm{L}^2 \Big(\frac{6.7 \times 10^{19}}{\alpha_{850}}\Big)\frac{\Gamma_0}{\Gamma_{RJ}} \ 10^{10} \mathrm{M}_\odot, 
\end{multline}

\noindent where $\nu_\mathrm{obs}$ = 345 GHz, $\nu_\mathrm{850}$ is the rest--frame frequency used to calibrate $\alpha_\mathrm{850}$, $D_\mathrm{L}$ is in Gpc and $\alpha_\mathrm{850}$ is the conversion factor between $L_\mathrm{870}$ and molecular gas mass. $\Gamma_\mathrm{RJ}$ is the correction for departures in the rest frame of the Planck function from Rayleigh–Jeans, with $\Gamma_\mathrm{0} (z=0, T_\mathrm{d}=25K) =$ 0.71 . We note that this method was calibrated using $\alpha_\mathrm{CO}$ = 6.5 M$_\odot$ (K km s$^{-1}$ pc$^2$)$^{-1}$, whereas we use $\alpha_\mathrm{CO}$ = 1 for our sample. Therefore, for a consistent comparison with CO(1--0), we divide the values obtained from the dust continuum by 6.5 to obtain total molecular gas masses. 

\begin{figure}
    \centering
    \hspace{-0.5cm}
    \includegraphics[width=\linewidth]{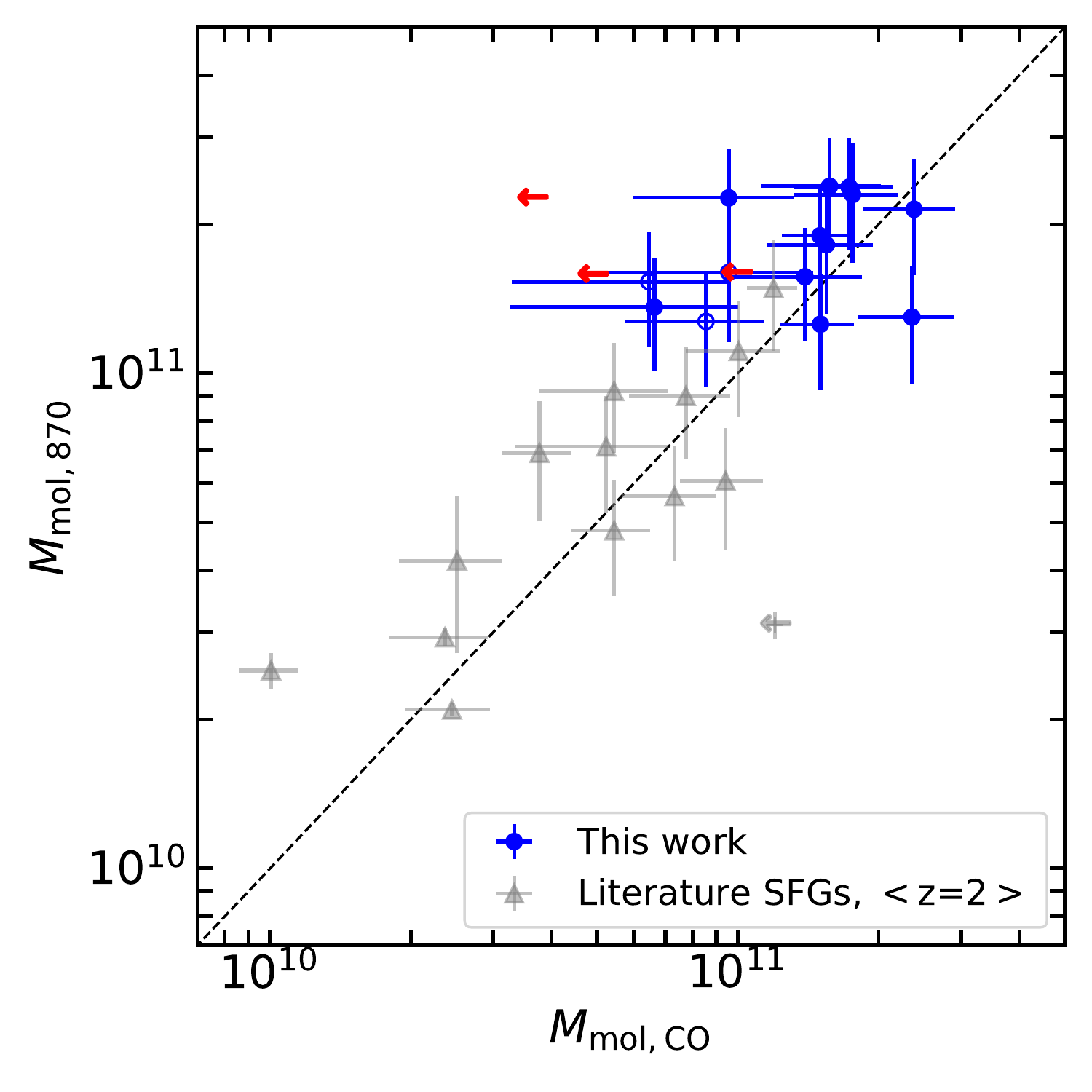}
    \caption{Comparison between the gas masses derived from CO(1--0) line luminosities and 870$\mu$m continuum flux densities for our sample, as well as with $z=2$ SFGs from \cite{boogaard2020,riechers2020a-vlaspecs} and \cite{kaasinen2019} with CO(1--0) line detections. We convert the molecular gas masses of the SFGs to the same $\alpha_\mathrm{CO}$ factor of 1 for an equal comparison with our sample. The gas masses agree within a factor of 2, although there is significant scatter around the 1:1 line shown by the dashed line.}
    \label{dust}
\end{figure}

Figure \ref{dust} shows the dust-based gas masses compared with the CO(1--0) gas masses. Since our sample probes a relatively narrow range in $S_\mathrm{850}$, we expand the range by comparing with $z\sim$ 2 SFGs from \cite{kaasinen2019} and the VLASPECS SFGs from \cite{boogaard2020} and \cite{riechers2020a-vlaspecs}. To test whether there is a correlation between the total molecular gas masses derived from CO(1--0) and 870 $\mu$m continuum, we calculate the Pearson's ($\tau$) and Spearman's ($\rho$) correlation coefficients. These coefficients measure the strength and direction of a linear and monotonic association (so we do not have to assume the underlying shape of the relation), respectively, between two variables. The coefficients can take values between +1 (perfect positive association) and $-$1 (perfect negative association), with 0 being no correlation. Assuming a null-hypothesis of no-correlation, the $p-$value represents the probability that the strength of the observed correlation is due to chance. We find $\tau$=0.1 ($p-$value = 0.7)  and $\rho$=0.2 ($p-$value=0.5) when considering just the values for our sample, meaning that the gas masses for our targets as traced by the two tracers do not appear to be correlated. We note however that we are limited by the small sample size. After including the VLASPECS \citep{riechers2020a-vlaspecs} and $z$=2 SFGs from \cite{kaasinen2019}, we find  $\tau$=0.9 ($p-$value = 4$\times$10$^{-8}$)  and $\rho$=0.8 ($p-$value = 3$\times$10$^{-7}$), showing a strong positive linear correlation between the two variables. 

We note that, at these redshifts, our continuum flux density measurements probe rest--frame wavelengths of $\sim250\mu$m, where the SED deviates significantly from the RJ tail of the dust emission. As a result, the 850$\mu$m continuum--based gas masses have considerable uncertainties. Nevertheless, they agree with the CO(1--0)--based gas masses to about a factor of two. However, caution is needed to draw strong conclusions from these results.

\subsection{CO Spectral Line Energy Distributions and Line Ratios} \label{sec:analysis-excitation}

\begin{figure}
    \hspace{-0.5cm}
    \includegraphics[width=\linewidth]{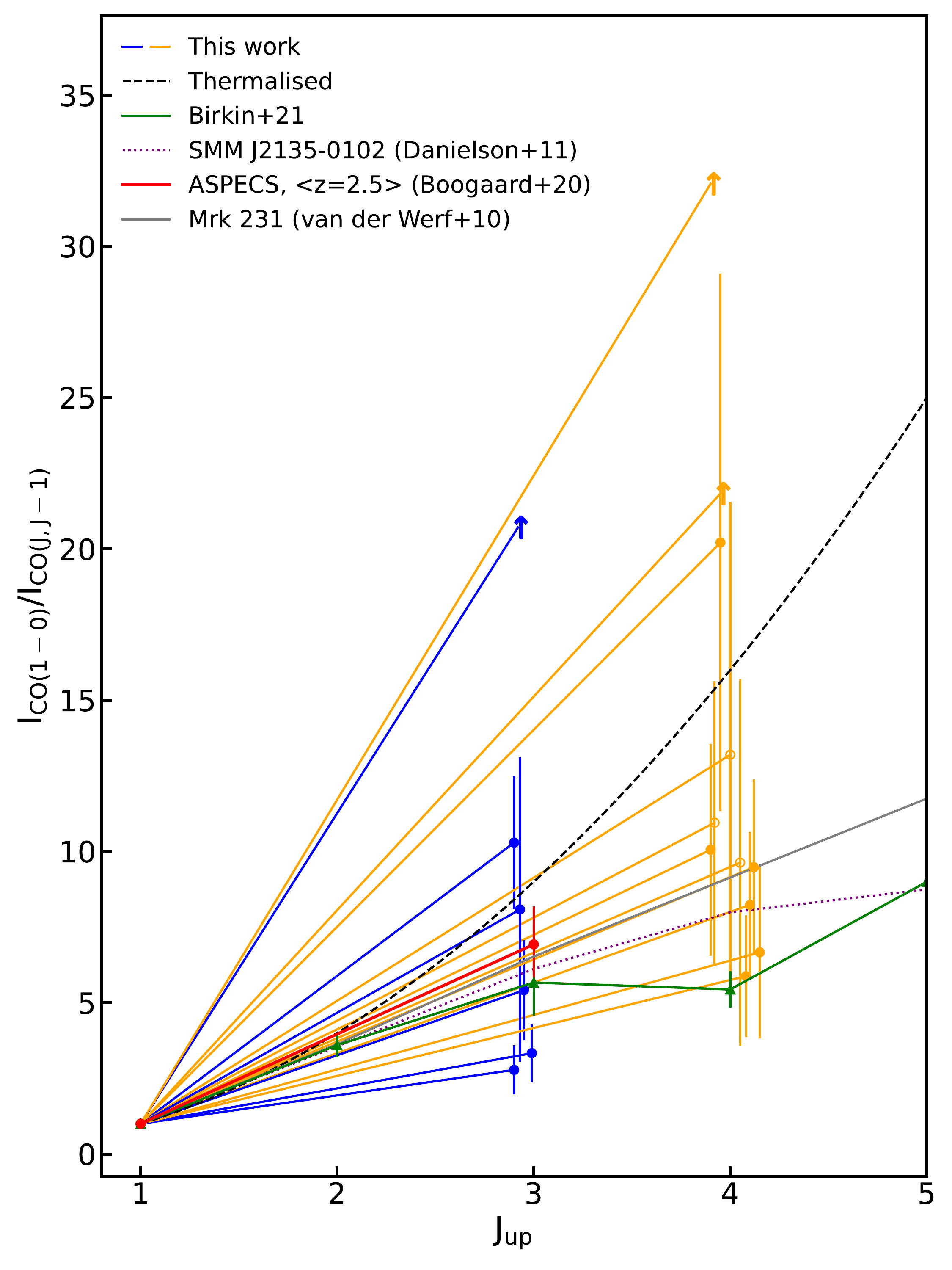}
    \caption{CO ladders for our targets normalised to the CO(1--0) integrated line flux. The targets show excitation that is, on average, comparable with the `Cosmic Eyelash' \citep[][]{danielson2011} and the average SMGs at high-redshift \citep[][]{birkin2020}, although there is a large scatter for both ratios. For comparison, we also show the average SLED for $z$=2.5 SFGs from the VLASPECS survey \citep[][]{boogaard2020,riechers2020a-vlaspecs} and the QSO Mrk 231 \citep{mrk231_cosled2010}.
    Data points are shifted on the $x$-axis for easier visualisation. }
    \label{sled}
\end{figure}

\begin{figure}
    \centering
    \includegraphics[scale=0.45]{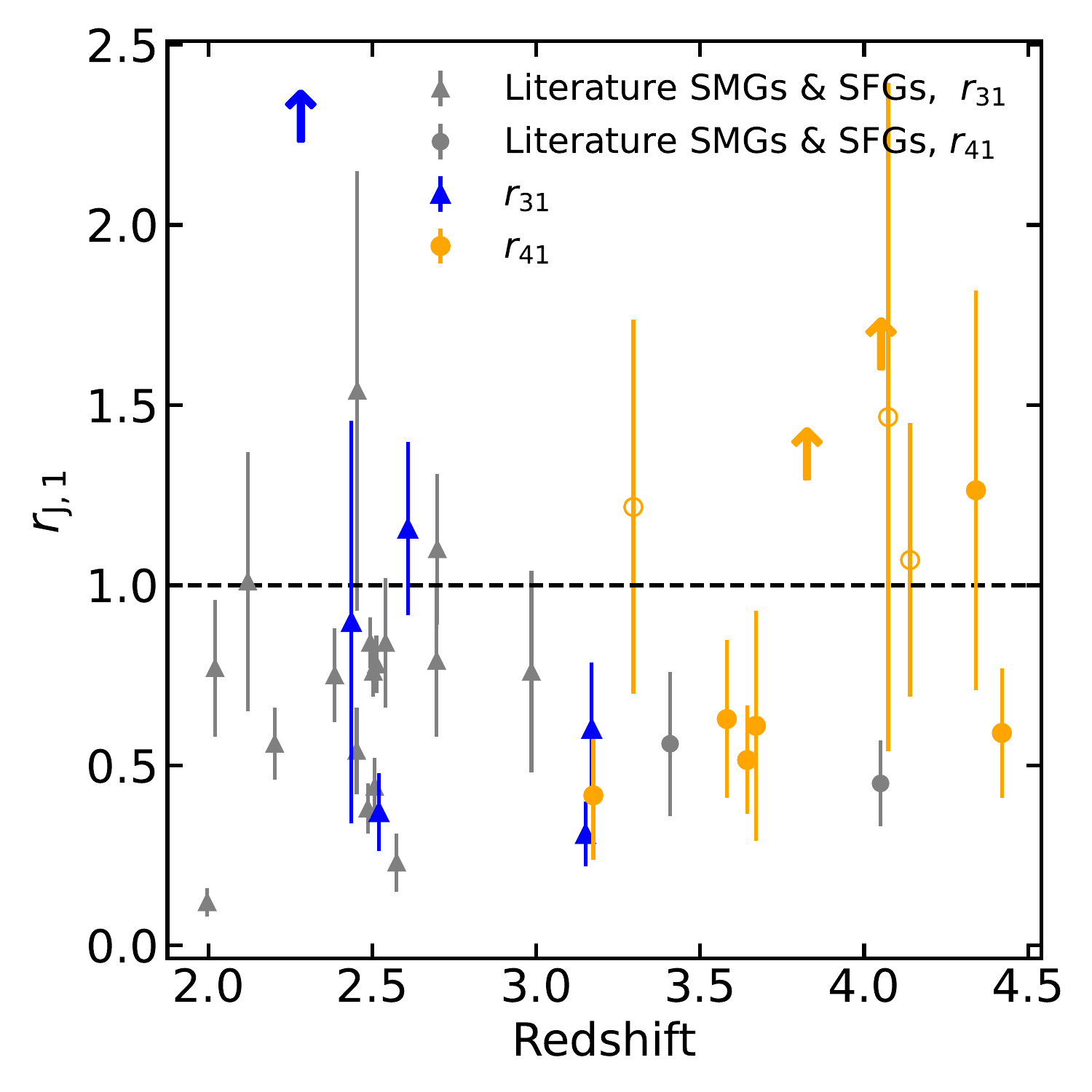}
    \includegraphics[scale=0.45]{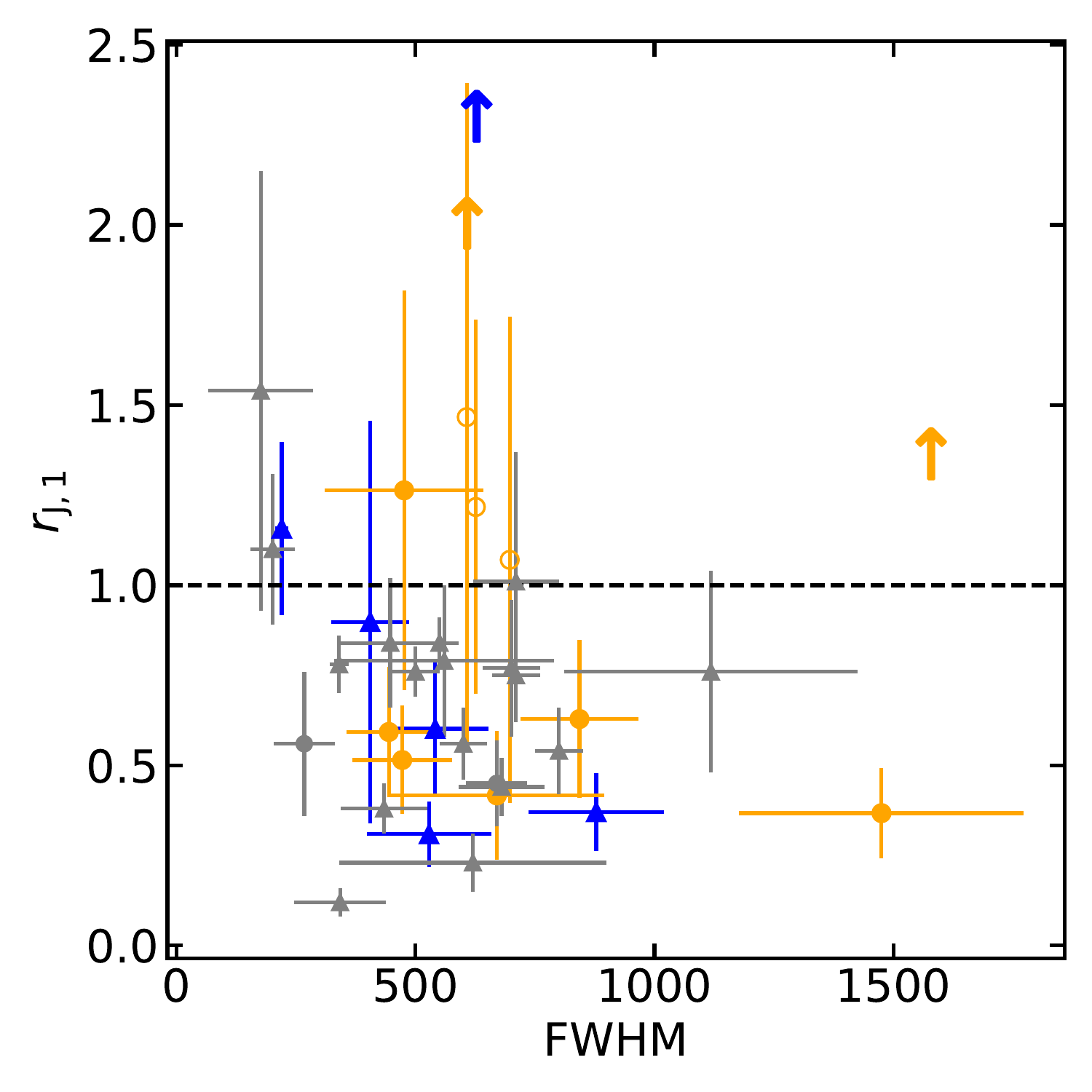}
    \includegraphics[scale=0.45]{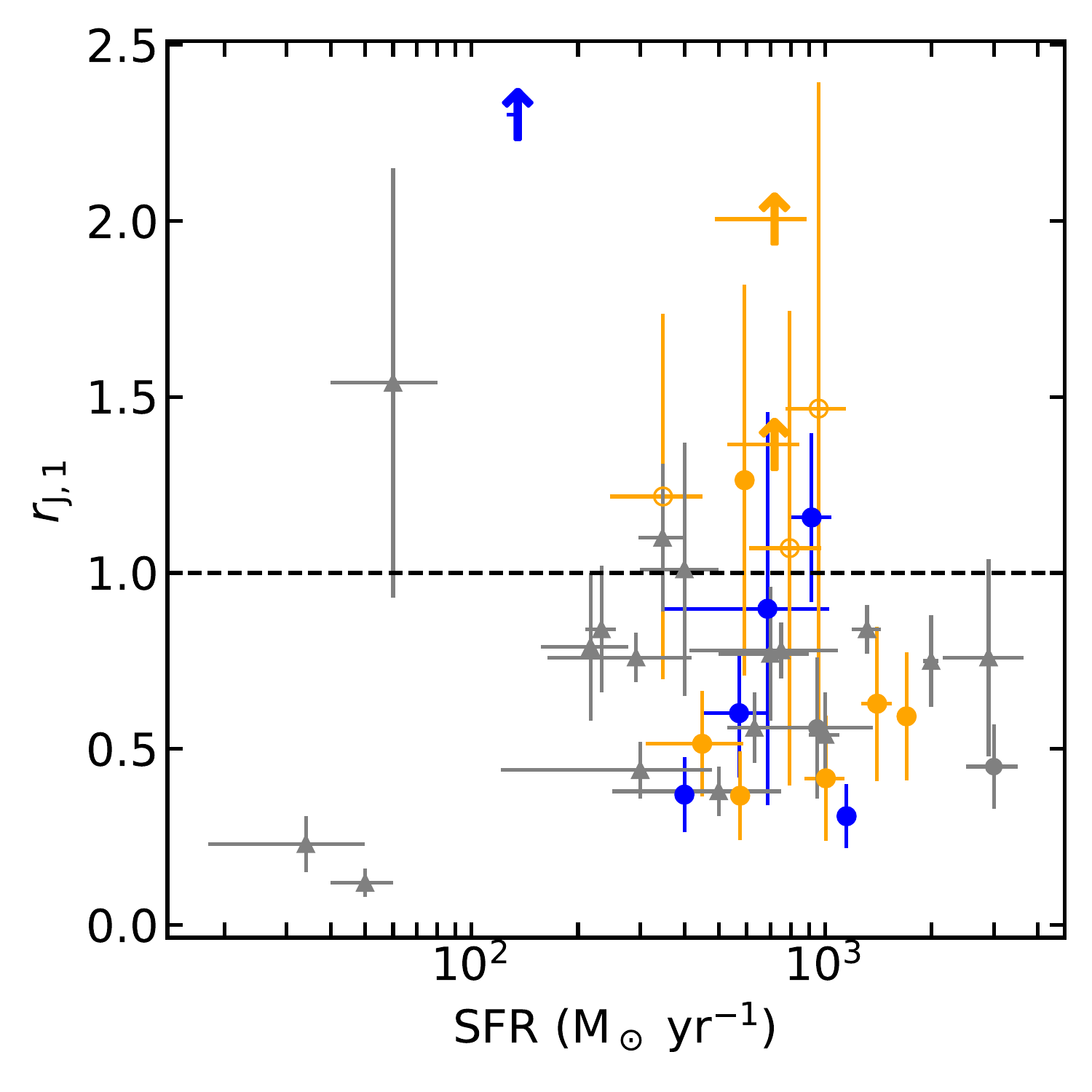}
    \caption{CO(3--2)/CO(1--0) (blue) and CO(4--3)/CO(1--0) (orange) excitation ratios as a function of redshift (top), CO(1--0) line width (middle) and SFR (bottom). Tentative detections are shown as open symbols, and 2$\sigma$ upper limits are marked by an upward pointing arrows. The dashed line shows the thermalised value $r_\mathrm{j1}$=1. We include literature values of unlensed SMGs and SFGs with CO(1--0) detections. We find no significant correlation for these variables with either ratio.}
    \label{ratios_z}
\end{figure}

The study of the CO excitation in galaxies provides key insights into the properties and state of their ISM, such as gas density and heating mechanisms. In Figure \ref{sled} we show the CO SLEDs for our sample normalised to the CO(1--0) transition. Because of the way our sample was constructed, each galaxy is only detected in either CO(3-2) or CO(4-3), which prevents modelling of the CO SLED for individual sources. However, we plot the statistical CO SLEDs derived by \citet{birkin2020} for $z$ = 1.2--4.8 SMGs and by \citet{boogaard2020} for $z$ = 2.5 SFGs, as well as for the ``Cosmic Eyelash'' \citep{danielson2011} for comparison. The non-detected sources, AS2COS0009.1, AS2UDS014.0 and AEG3, show super-thermal excitation, as their CO(3--2) or CO(4--3) luminosity is significantly higher than our 2$\sigma$ upper limits on CO(1--0). All the other galaxies in our sample are consistent with thermal or sub-thermal excitation within the error bars. The $I_\mathrm{CO(3-2)}$/$I_\mathrm{CO(1-0)}$ excitation is also consistent with the \cite{boogaard2020} and \cite{birkin2020} statistical SLEDs, while $I_\mathrm{CO(4-3)}$/$I_\mathrm{CO(1-0)}$ appears to be slightly higher than reported in \cite{birkin2020}.

The ratios of CO line luminosities can be used as excitation indicators of the average state of the molecular gas, and are usually reported as:

\begin{equation}
r_{J1} = \frac{L'_\mathrm{CO(J-(J-1))}}{L'_\mathrm{CO(1-0)}}.    
\end{equation}

We measure line luminosity ratios between the CO(3--2) and the CO(1--0) emission lines of 0.31--1.16, with a median of $r_{31}$ = 0.75$\pm$0.39. For the CO(4--3) line, we find ratios in the range 0.37--1.26, with a median $r_{41}$ = 0.63$\pm$0.44, where the errors are calculated as the median absolute deviation. These ratios, listed in Table \ref{tab:ratios}, reveal a large spread in the excitation conditions from galaxy to galaxy, from low excitation ($r_{31} \sim$ 0.3) to super-unity ratios ($r_{31} \sim$ 1.2, $r_{41} \sim$ 1.5). 
Under normal conditions in SMGs, the ISM is dominated by cold molecular gas and the CO emission is optically thick, leading to $r_{J1} \leq$1. The super-thermal CO excitation can occur if a) the CO emission is optically thin, b) CO(1--0) is self-absorbed, c) emission is coming from ensembles of small, unresolved, optically thick clouds or d) the emission is optically thick and has temperature gradients \citep{bolatto2000,bolatto2003}.
It is also possible that, due to the low SNR of our detections, we are missing low surface brightness CO(1--0) emission below our detection threshold. 

\begin{table}
\caption{CO line ratios \label{tab:ratios} }
\hspace{-0.5cm}
\begin{tabular}{@{}lccccc@{}}
 \hline \hline
Target & $r_{31}$ & $r_{41}$ \\
\hline
AS2COS0008.1 & -- & 0.63$\pm$0.22 \\
AS2COS0009.1 & $>$0.97 & -- \\
AS2COS0013.1 & 1.16$\pm$0.24 & -- \\
AS2COS0023.1 & -- & 1.26$\pm$0.56 \\
AS2COS0031.1 & -- & 0.51$\pm$0.15  \\
AS2COS0054.1 & -- & 0.42$\pm$0.18 \\
AS2UDS010.0 & 0.60$\pm$0.18 & -- \\
AS2UDS011.0$^\ast$ & -- & 1.46$\pm$0.93 \\
AS2UDS012.0 & 0.37$\pm$0.11 & -- \\
AS2UDS014.0 & -- & $>$0.68 \\
AS2UDS026.0$^\ast$ & -- & 1.22$\pm$0.51 \\
AS2UDS126.0 & 0.90$\pm$0.56 & \\
AEG2 & -- & 0.61$\pm$0.32 \\
AEG3 & -- & $>$1.67 \\
CDFN1 & 0.31$\pm$0.09 & -- \\
CDFN2 & -- & 0.59$\pm$0.18 \\
CDFN8 & -- & 1.07$\pm$0.38  \\
\hline
Median & 0.75$\pm$0.39 & 0.63$\pm$0.44 \\
\hline
  \hline
 \end{tabular}
\end{table}

The median $r_{31}$ value is comparable to that of other high-redshift galaxies reported in the literature. In the VLASPECS survey \citep{riechers2020a-vlaspecs}, the median $r_{31}$ for MS galaxies at $z$ = 2--3  is 0.84$\pm$0.26; \cite{xiao2022} reported an $r_{31}$ of 0.8 for two starburst galaxies in a $z$ = 2.5 protocluster, and \cite{sharon2016} found $r_{31}$ = 0.78 $\pm$ 0.27 for a sample of lensed and unlensed SMGs at $z \sim$ 2. Our value for $r_{31}$ is higher than the ratio found by \cite{ivison2011}, 0.55$\pm$0.05, for SMGs at $z = 2.2-2.5$, and consistent with 0.63$\pm$0.12 from \cite{birkin2020}. There are fewer data points available on $r_{41}$: our $r_{41}$ = 0.63$\pm$0.11 is consistent with $r_{41}$=0.56$\pm$0.20 and 0.45$\pm$0.12, found respectively, for the unlensed SMGs J13120+4242 and GN20 \citep{friascastillo2022,carilli2010gn20sled}.

\begin{figure*}
    \centering
    \includegraphics[scale=0.5]{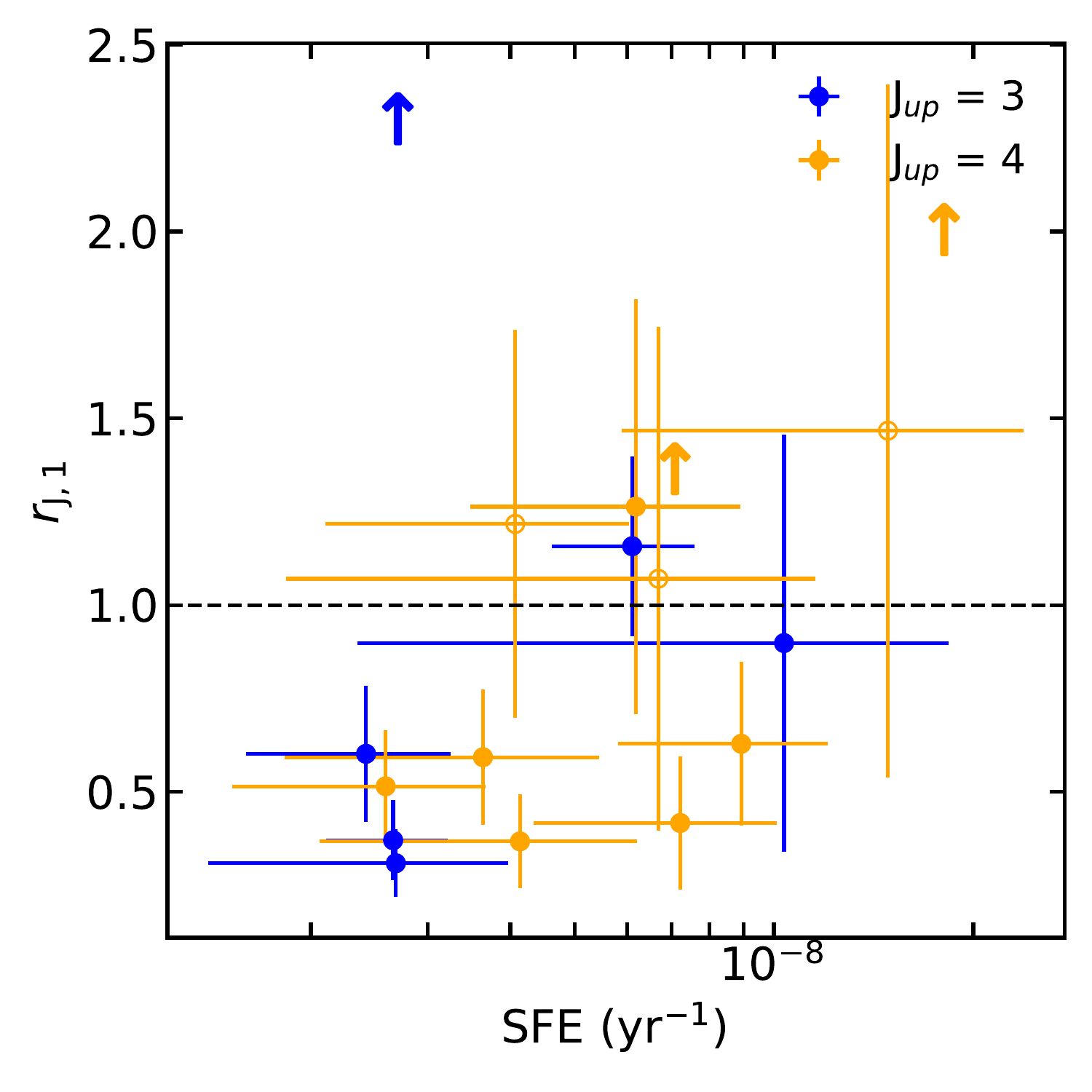}
    \includegraphics[scale=0.54]{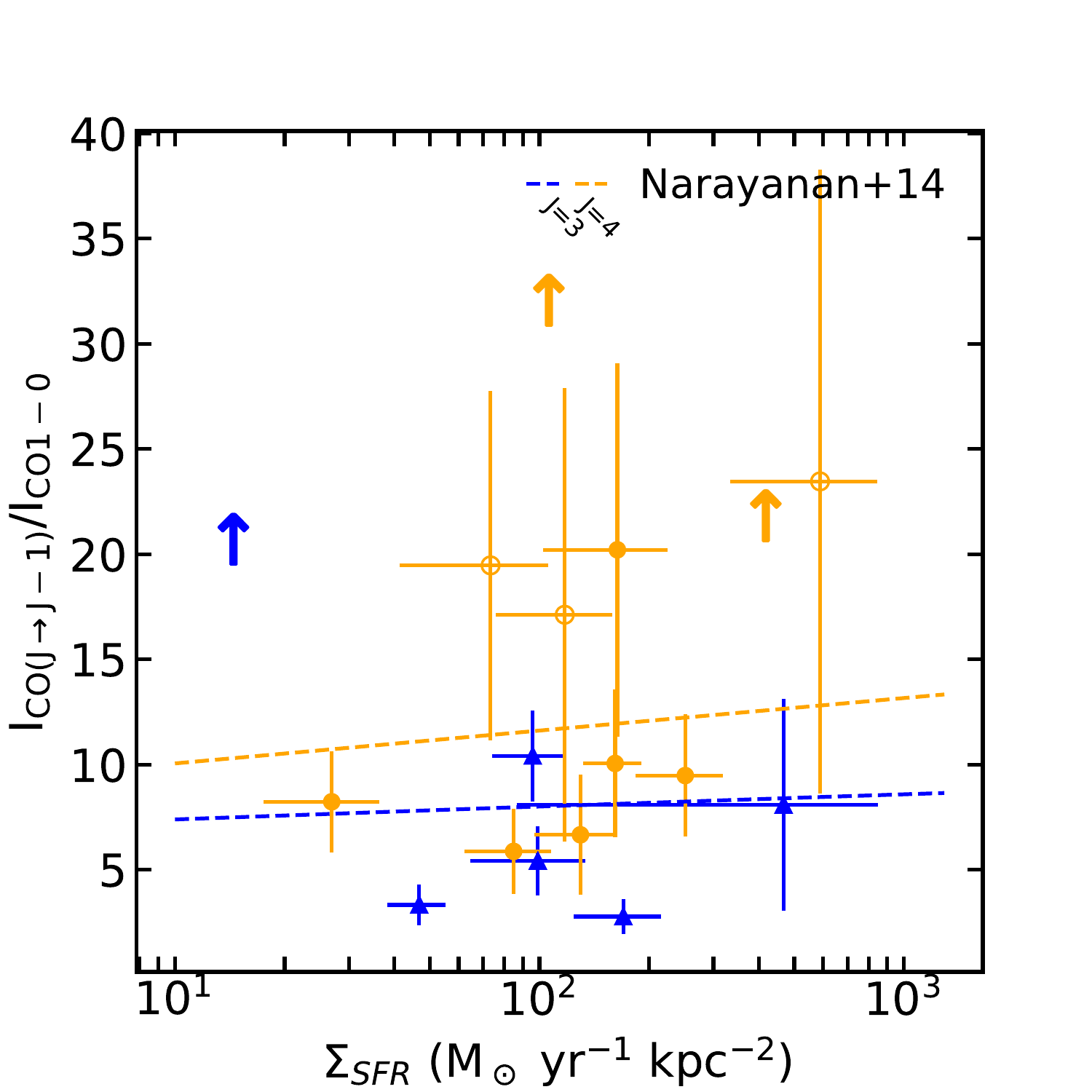}
    \caption{CO(3--2)/CO(1--0) (blue) and CO(4--3)/CO(1--0) (orange) excitation ratios as a function of star-formation efficiency (left) and star-formation rate surface density (right). Open symbols indicate tentative detections, while upward pointing arrows indicate 2$\sigma$ upper limits. We calculate Pearson’s and Spearman’s correlation coefficients to look for monotonic trends for these variables and find no significant correlation with SFE. The Spearman's test does reveal a positive correlation of $r_{41}$ with $\Sigma_\mathrm{SFR}$, with a $p-$value of 0.02, although we are limited by the small sample size and the lack of literature data. Compared to the line excitation predictions from \cite{narayanan_krumholz2014} (dashed lines), our sources show a large amount of scatter around the models. }
    \label{ratios_sfe}
\end{figure*}

We now look for correlations between the molecular gas excitation and other observed and derived quantities. The 870 $\mu$m flux density is derived from ALMA measurements \citep[][Chapman et al., in prep]{stach2019,simpson2020}, and we calculate the MS offset following \cite{speagle2014}. Although recent studies found evidence for a flattening of the MS at the high-stellar-mass end \citep{whitaker2014,schreiber2015,leslie2020}, this is mostly seen at $z <$ 1.5. At high-redshift, a linear model is still preferred by the available data \citep{lee2018,leslie2020}. SFR and stellar masses are obtained through SED fitting as described in Section \ref{sec:Results}. We calculate the Spearman's and Pearson's rank correlation coefficients for both ratios separately (including the literature compilation) and find no significant trends with redshift, CO(1--0) FWHM, SFR (Figure \ref{ratios_z}), SFR surface density (Figure \ref{ratios_sfe}), 870$\mu$m flux density and MS offset (not pictured).

A previous study by \cite{sharon2016} has found a positive trend between star formation efficiency (SFE = SFR/$M_\mathrm{mol}$) and CO excitation in a sample of SMGs and AGN at $z \sim$ 2--3. Similar trends have also been found in the local Universe, in ULIRGs \citep{papadopoulos2012,greve2014} and IR-luminous galaxies \citep{yao2003}. We show our ratios as a function of SFE in Figure \ref{ratios_sfe} (left). We calculate the Spearman's and Pearson's rank correlation coefficients for both ratios together and independently (assuming non-detections as lower limits), but do not find any significant correlation between line excitation and SFE. 

The CO excitation has also been proposed to correlate with the SFR surface density, $\Sigma_\mathrm{SFR}$ \citep{narayanan_krumholz2014,daddi2015,boogaard2020,valentino2020}. Based on high-resolution ALMA imaging, the dust-emitting regions of our targets in the COSMOS and UDS fields have radii in the range 0.7 -- 2.4 kpc, with a median radius of 1.4$\pm$0.5 kpc \citep[Ikarashi et al. 2023, in prep]{gullberg2019,stach2019,simpson2020}. The SMA 850$\mu$m imaging of the sources in the EGS and GOODS-North fields does not have the necessary resolution to derive dust continuum sizes, so we use the median size of the AS2COSMOS and AS2UDS sources, 1.4$\pm$0.5 kpc. We calculate $\Sigma_\mathrm{SFR}$, and plot it as a function the flux ratios of CO(3--2) and CO(4--3) over CO(1--0) in Figure \ref{ratios_sfe} (right). We note that the dust continuum traces star-forming regions, and is often found to be smaller than the extent of the CO(1--0)-traced cold molecular gas reservoirs \citep[e.g.,][]{simpson2015a,barro2016,hodge2016,dannerbauer2017,calistro-rivera2018,simpson2020}. 

To assess the correlation between the CO excitation and $\Sigma_\mathrm{SFR}$, we again calculate the corresponding Spearman's and Pearson's rank correlation coefficients. For $r_{31}$, the tests find no significant correlation (Spearman's $\rho$=-0.3, $p$=0.6 and Pearson's $\tau$=0.2, $p$=0.7). For $r_\mathrm{41}$, the Spearman's test shows a strong positive correlation between the variables, $\rho$=0.8, and a $p-$value of 0.023, meaning that we can reject the null hypothesis that the samples are uncorrelated, while Pearson's test finds a strong positive correlation ($\tau$=0.7) but we cannot rule out the possibility of them being uncorrelated from the $p$-value of 0.1. Since we are limited by the small sample size and large error bars, we refrain from fitting the data. We compare these to the line excitation predictions from \cite{narayanan_krumholz2014} for unresolved observations of CO transitions, shown by the dashed lines for the two ratios $r_{31}$ and $r_{41}$ in blue and orange, respectively. While the ratios qualitatively agree with the theoretical models, observations show a large amount of scatter.

\subsection{Comparison With Semi-analytic Models}

How do our measurements of molecular gas mass in bright SMGs compare to current theoretical models? Due to their $M_\star$ and SFR, SMGs pose a challenge to galaxy evolution simulations. Recently, state-of-the-art hydrodynamical simulations such as EAGLE, SIMBA, and Illustris TNG have been able to reproduce some key aspects of the SMG population, albeit using very different assumptions. However, due to their limited simulation volumes, they do not contain enough sources with $S_\mathrm{850}$ $\geq$10~mJy, whereas our sample spans $S_\mathrm{850}\approx10-20$~mJy. 

We have chosen to compare our data to predictions from the SHARK semi-analytical model (SAM) \citep{lagos2018} applied to the \textsc{Surfs} N-body simulation \citep{Elahi2018}. SHARK has been able to reproduce SMG number counts and redshift distribution without having to introduce e.g. top-heavy stellar initial mass function \citep{lagos2020}. We adopt the broadband continuum fluxes predicted using the \citet{Lagos2019} framework. Thanks to the large box size of the \textsc{Surfs} simulation (210$^3$~(cMpc/$h$)$^3$), SHARK includes a sizeable population of $S_\mathrm{850}\geq$10~mJy SMGs which can be directly compared to our sample. Such bright SMGs are mostly absent from the recent hydrodynamical simulations such as EAGLE \citep{schaye2015, mcalpine2019} or SIMBA \citep{Dave2019, Lovell2021}, as these have volumes $\approx$10$\times$ smaller.

\begin{figure}
    \centering
    
    \includegraphics[width=0.99\linewidth]{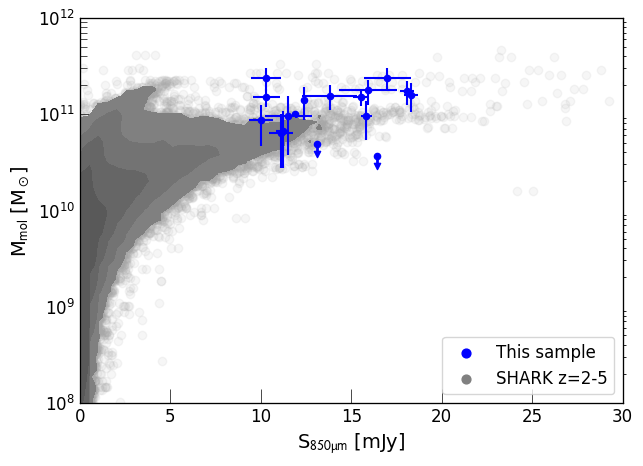}
    \caption{Comparison of 850$\mu$m flux density and $M_\mathrm{mol}$ between our sample (blue) and $z=2-5$ galaxies from the SHARK semi-analytic model (gray). We find reasonable agreement between SHARK predictions and our observations (assuming $\alpha_\mathrm{CO}=1.0)$.}
    \label{fig:shark_850um_Mmol}
\end{figure}

\begin{figure*}
    \centering
    
    \includegraphics[height=8cm]{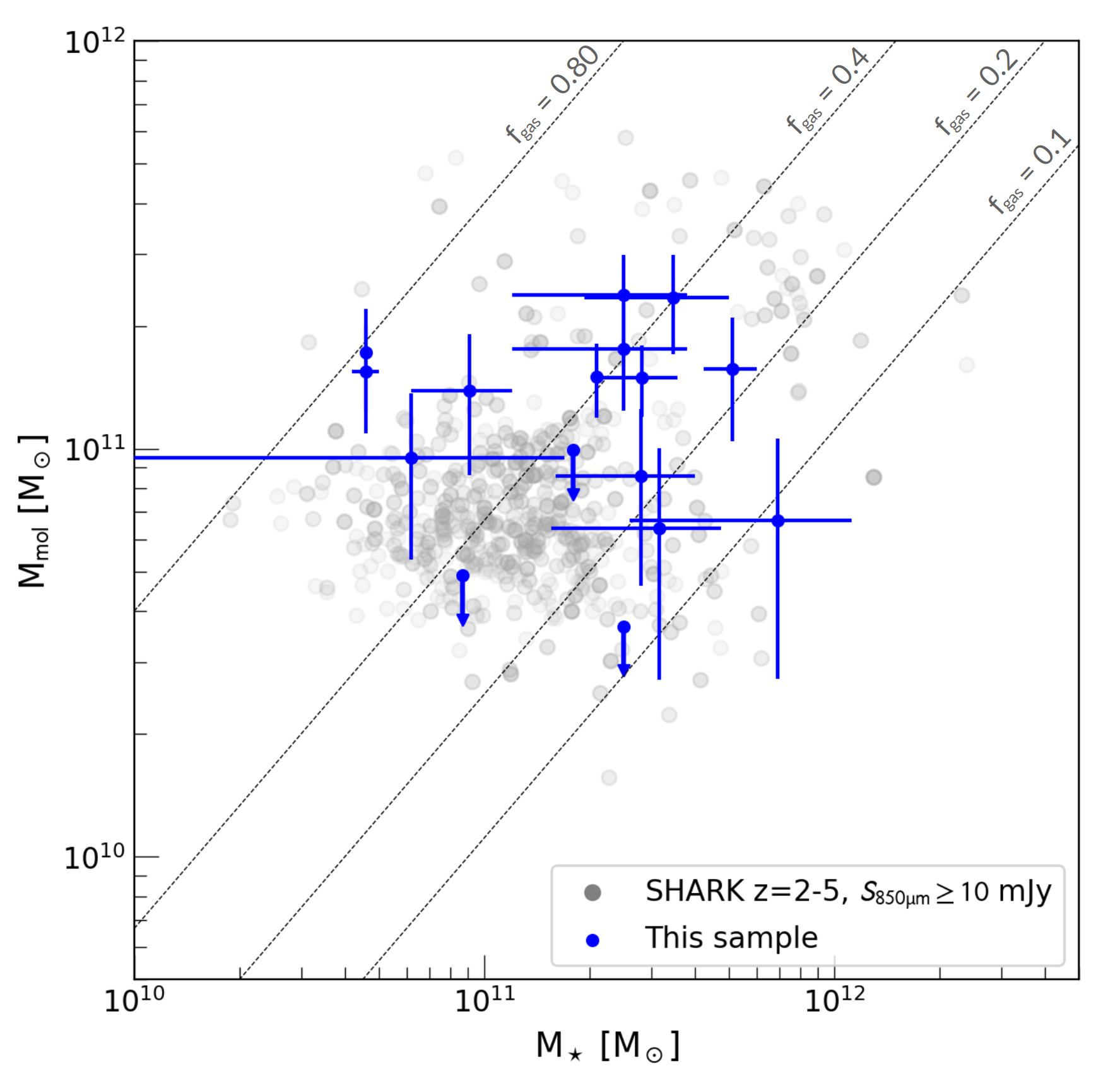}
    \includegraphics[height=8cm]{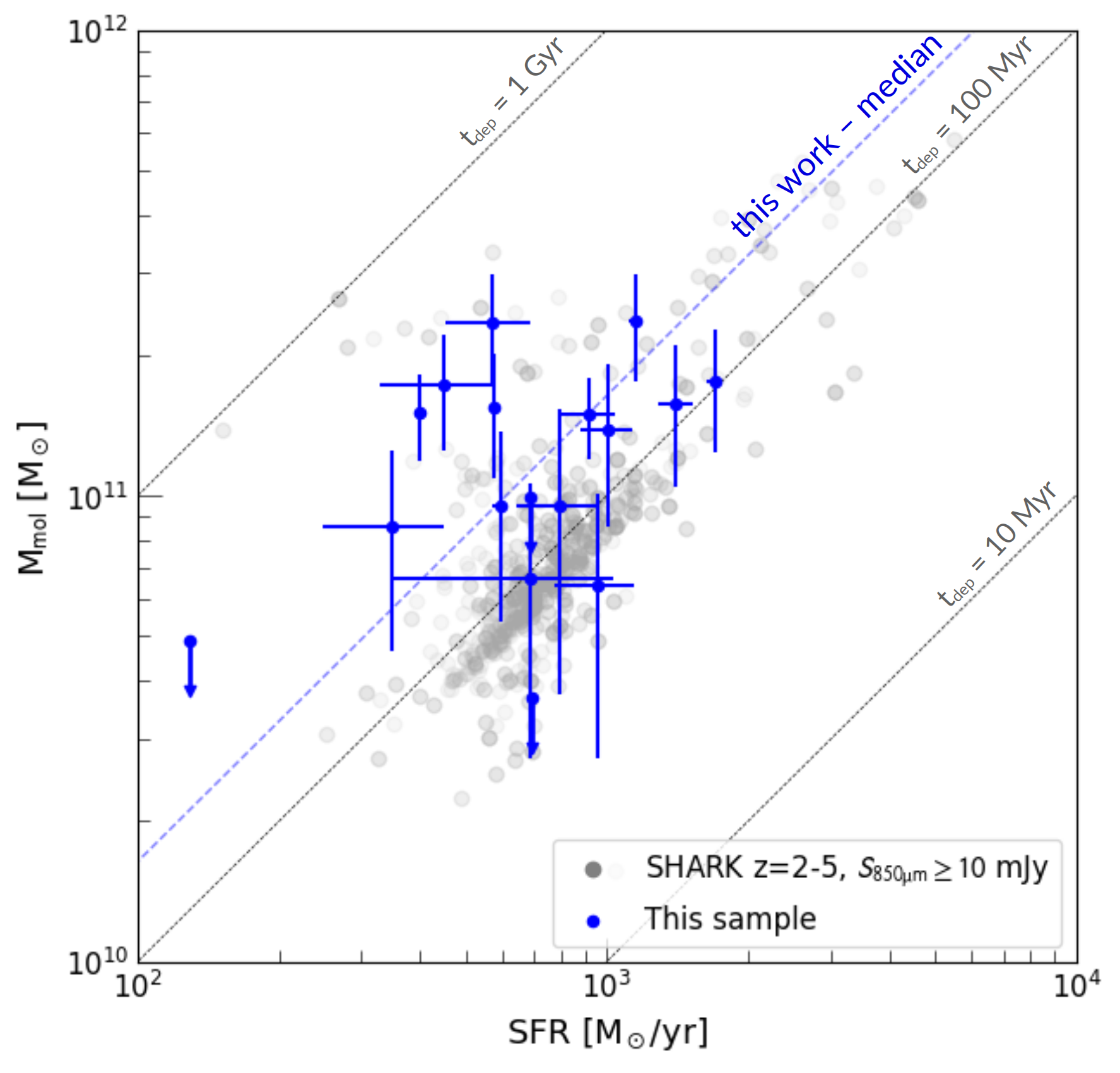}
    \caption{Comparison of our sample with SHARK $z=2-5$ SMGs with $S_\mathrm{850}\geq$10~mJy. In the $M_\star$--$M_\mathrm{mol}$ plane (\textit{left}), our observations yield $M_\mathrm{mol}$ on the upper end of the SHARK distribution. The SFR--$M_\mathrm{mol}$ plane (\textit{right}) reveals a clear discrepancy in molecular gas depletion timescale $t_\mathrm{dep}$: while SHARK assumes a fixed $t_\mathrm{dep}$=100 Myr in the ``starburst'' mode (notice the conspicuous diagonal trend in the model points), we find a median $t_\mathrm{dep}$=140~Myr. Despite this limitation, we find a good agreement with our observations, which ishow a median of $\approx140$~Myr}
    \label{fig:shark_Mstar_Mmol}
\end{figure*}

For comparison, we select galaxies from the SHARK catalogue spanning $z=2-5$. We compare four key parameters: $S_\mathrm{850}$, molecular gas mass $M_\mathrm{mol}$, $M_\star$, and SFR. As shown in Fig.~\ref{fig:shark_850um_Mmol}, SHARK galaxies with $S_\mathrm{850}\geq$10~mJy have molecular gas masses consistent with our measurements. At higher 850$\mu$m fluxes, the observed $M_\mathrm{mol}$ values tend to fall on the upper envelope of SHARK predictions. This might be a result of an observational bias: SMGs with lower $f_\mathrm{gas}$ will be fainter in mid-$J$ CO emission and thus more difficult to confirm spectroscopically. However, we note that the spectral survey of \citet{birkin2020} shows $\sim$75\% detection rate; it is thus unlikely that we are missing a substantial population of 850$\mu$m--bright, gas-poor SMGs. 
We note that our choice of $\alpha_\mathrm{CO}$=1 for converting the CO(1--0) luminosity to $M_\mathrm{mol}$ facilitates the agreement between SHARK predictions and our observations: for $\alpha_\mathrm{CO}\approx$4, SHARK would have underestimated $M_\mathrm{mol}$ by a factor of a few.

Fig.~\ref{fig:shark_Mstar_Mmol} shows a more detailed comparison with $z=2-5$ SHARK SMGs with $S_\mathrm{850}\geq$10~mJy; the SHARK sample totals 1152 galaxies with a median redshift of 2.53. In the $M_\star$--$M_\mathrm{mol}$ plane, we see a generally good agreement between gas fractions of SHARK SMGs and our sample. Similar to Fig.~\ref{fig:shark_850um_Mmol}, at $M_\star\leq10^{11}$~$M_\odot$, our observations probe the upper end of molecular gas masses predicted by SHARK. 
Finally, we compare our data and SHARK SMGs in the SFR--$M_\mathrm{mol}$ plane (Fig.~\ref{fig:shark_Mstar_Mmol}, right), although there are systematic differences due to the assumption made in SHARK. Namely, SHARK assumes two modes of star-formation: a ``normal' mode with $t_\mathrm{dep}$=1~Gyr, and a ``burst'' mode with $t_\mathrm{dep}$=100~Myr. Despite this limitation, we find a good agreement with our observations which imply $t_\mathrm{dep}\approx50-400$~Myr, with a median of $\approx140$~Myr (see Section 4.2). 

The good agreement between our sample and its SHARK counterparts highlights the power of the state-of-art simulations to reproduce the bulk properties of the SMG population. In particular, as the SHARK model is not tuned to reproduce high-redshift galaxy population, the fact that it reproduces the key properties of the massive end of the SMG population is a remarkable achievement.

\section{Conclusions} \label{sec:conclusions}

We have presented new, deep VLA observations of the CO(1--0) emission in 17 dusty star-forming galaxies at $z$ = 2--4 from the on-going VLA Legacy Survey of Molecular Gas at High Redshift. We have successfully detected CO(1--0) in 11 sources, with three further tentative detections. These systems are representative of the most massive, star-forming galaxies at their redshift. Our main findings are as follows:\\

    $\bullet$  Our galaxies have total molecular gas masses in the range 6--23 $\times$ 10$^{10}$ M$_\odot$, assuming a CO-to-H$_2$ conversion factor of 1. Combined with stellar masses and SFRs obtained via SED fitting, we find a median $f_\mathrm{gas}$ of 0.35$\pm$0.21 and a median $t_\mathrm{dep}$ of 140$\pm$70 Myr. These values are below what would be expected from empirical scaling relations, but are consistent with median values found in large surveys of SMGs with  mid--$J$ CO observations. We do not see evidence of evolution of $f_\mathrm{gas}$ or $t_\mathrm{dep}$ with redshift.
        
    $\bullet$ Combining our data with ALMA and NOEMA CO($J$=3--2) and CO($J$=4--3) observations yield median excitation ratios of $r_{31}$ = 0.75$\pm$0.39 and $r_{41}$ = 0.63$\pm$0.44, although there is significant scatter within the sample.

    $\bullet$ We supplement our sample with available literature data of unlensed SMGs, and quantitatively investigate the correlation of the excitation ratios with a number of parameters (e.g., $z$, CO(1--0) line width, SFE, $\Sigma_\mathrm{SFR}$). We find no significant trends with any of these parameters, except for a tentative positive correlation between $r_{41}$ and $\Sigma_\mathrm{SFR}$ for our sources, although we are limited by the sample size. A more in-depth analysis will be available when the full sample has been observed.
    
    $\bullet$ Finally, we compare our data with the population of $S_{850}>$10 mJy SMGs from the SHARK catalogues \citep{lagos2018}. The gas fractions and depletion times of our source show a remarkably good agreement with those of their SHARK counterparts, highlighting the power of current state-of-the-art simulations and their potential synergy with future CO(1--0) surveys of high--redshift galaxies.

Our results highlight the heterogeneous nature of the most massive, star-forming galaxies at $z\sim$2.5--4, and the importance of CO(1--0) observations to robustly constrain their ISM properties. Almost 60\% completed, the VLA Legacy Survey of Molecular Gas at High Redshift already provides the basis for follow-up studies, such as the high-resolution VLA CO(1--0) imaging
necessary for detailed morphological and dynamical studies, which are currently limited to a handful of the most extreme cases \citep[e.g.,][]{hodge2012,friascastillo2022}.

\begin{table*}
\caption{List of high-redshift non-lensed SMGs with CO(1--0) detections and available parameters used in the figures in this work. References are as follows: [1] \citet{huynh2017}, [2] \citet{calistro-rivera2018},[3] \citet{dacunha2015}, [4] \citet{ivison2011}, [5] \citet{tacconi2006}, [6] \citet{riechers2011b},[7] \citet{thompson2012}, [8] \citet{weiss2009}, [9] \citet{xiao2022}, [10] \citet{leung2019}, [11] \citet{greve2005}, [12] \citet{bothwell2013}, [13] \citet{sharon2016}, [14] \citet{friascastillo2022}, [15] \citet{carilli2010gn20sled}, [16] \citet{daddi2009a} \label{tab:lit} }
\begin{center}
\begin{adjustwidth}{-2.9cm}{}
 \begin{tabular}{@{}lcccccccccc@{}}
 \hline
Name & $z$ & $L'_\mathrm{CO(1-0)}$ ($\times$10$^{10}$) & FWHM$_\mathrm{CO(1-0)}$ & $r_{31}$ & $r_{41}$ & SFR & $M_*$ ($\times$10$^{10}$) & $f_\mathrm{gas}$ & $t_\mathrm{dep}$ & Ref.\\
 &  & [K km s$^{-1}$ pc$^{-2}$] & [km s$^{-1}$] & & & [M$_\odot$ yr$^{-1}$] & [M$_\odot$] & & [Myr] & \\
\hline
ALESS122.2 & 2.02&13$\pm$2 & 700$\pm$60 & 0.77$\pm$0.19 & --& 700$\pm$200 & 8$\pm$5 & 0.62$\pm$&  140$\pm$& [1],[2],[3]\\
ALESS67.1 &2.12 &10$\pm$2 & 710$\pm$90 & 1.01$\pm$0.36 &-- & 400$\pm$100 & 24$\pm$21 &0.29$\pm$ & 130$\pm$ & [2],[3]\\
J123549+6215 &2.202 &8$\pm$1 & 600$\pm$50 & 0.56$\pm$0.1 &-- &630$\pm$100 & 21$\pm$6 & 0.27$\pm$& 123$\pm$ & [4],[5]\\
J16350+4057 & 2.385& 8$\pm$1& 710$\pm$50 & 0.75$\pm$0.13 &-- & 1995$\pm$100 & 14$\pm$4 & 0.37$\pm$& 41$\pm$ & [4],[5] \\
J123707+6214 & 2.452& 11$\pm$2& 800$\pm$50 & 0.54$\pm$0.12 &-- & 1000$\pm$100 & 13$\pm$3 & 0.45$\pm$& 108$\pm$ & [4],[6] \\
J14009+0252 & 2.486&  10$\pm$1& 434$\pm$90 & 0.38$\pm$0.07 &-- & 500$\pm$250 & 16$\pm$4 & 0.38$\pm$& 160$\pm$& [7],[8] \\
SB1 & 2.494 & 4.9$\pm$0.4  & 550$\pm$40 & 0.84$\pm$0.07 &-- & 1300$\pm$120 &9$\pm$2 &0.70$\pm$0.1 & 150$\pm$20 &  [9] \\
MS1 & 2.503 & 2.3$\pm$0.2 & 500$\pm$50 & 0.76$\pm$0.07 & -- & 290$\pm$128 & 23$\pm$2& 0.29$\pm$0.1 & 320$\pm$140 & [9] \\
MS2 & 2.507 & 1.8$\pm$0.3 & 680$\pm$90 & 0.44$\pm$0.08 &-- & 300$\pm$179 & 22$\pm$2& 0.25$\pm$0.1 & 250$\pm$150 & [9]\\
SB2 & 2.512 & 3.2$\pm$0.3 & 340$\pm$20 & 0.78$\pm$0.08 &-- & 750$\pm$340 & 7$\pm$2 &0.66$\pm$0.1 & 180$\pm$80 & [9] \\
HXMM05 & 2.985 & 27$\pm$9 & 1100$\pm$300 & 0.76$\pm$0.28 &-- & 2900$\pm$750 & 12$\pm$7 & 0.2$\pm$0.2 & 72$\pm$27 & [10] \\
J22174+0015 & 3.099 & 4$\pm$1 & 560$\pm$110 & 0.79$\pm$0.29 &-- & -- & -- & --& --& [11],[12],[13] \\
J13120+4242& 3.408& 10$\pm$3 & 267$\pm$64 & -- & 0.56$\pm$0.20 &950$\pm$420 & 65$\pm$25 &0.17$\pm$0.03 & 137$\pm$69 & [14],[11] \\
GN20 & 4.05 &16$\pm$0.4 & 670 & -- & 0.45$\pm$0.12 & 3000 & 23 & 0.4 & 50 & [15],[16] \\
\hline
 \end{tabular}
 \end{adjustwidth}
 \end{center}
\end{table*}

\acknowledgements{
The National Radio Astronomy Observatory is a facility of the National Science Foundation operated under cooperative agreement by Associated Universities, Inc.\\
M.F.C. and J.A.H. acknowledge support of the VIDI research programme with project number 639.042.611, which is (partly) financed by the Netherlands Organisation for Scientific Research (NWO). M.R. is supported by the NWO Veni project ``Under the lens" (VI.Veni.202.225). I.R.S., J.E.B. and A.M.S. acknowledge support from STFC (ST/T000244/1). JEB acknowledges the support
of STFC studentship (ST/S50536/1). C.-C.C. and C.-L.L. acknowledge support from the National Science and Technology Council  of Taiwan (NSTC 109-2112-M-001-016-MY3  and 111-2112-M-001-045-MY3), as well as Academia Sinica through the Career Development Award (AS-CDA-112-M02). Funded by the Deutsche Forschungsgemeinschaft (DFG, German Research
Foundation) under Germany's Excellence Strategy -- EXC-2094 --
390783311.}

\appendix
\addcontentsline{toc}{section}{Appendices}

\subsection*{Curve of Growth}

\begin{figure*}
    \centering
    \includegraphics[width=\textwidth]{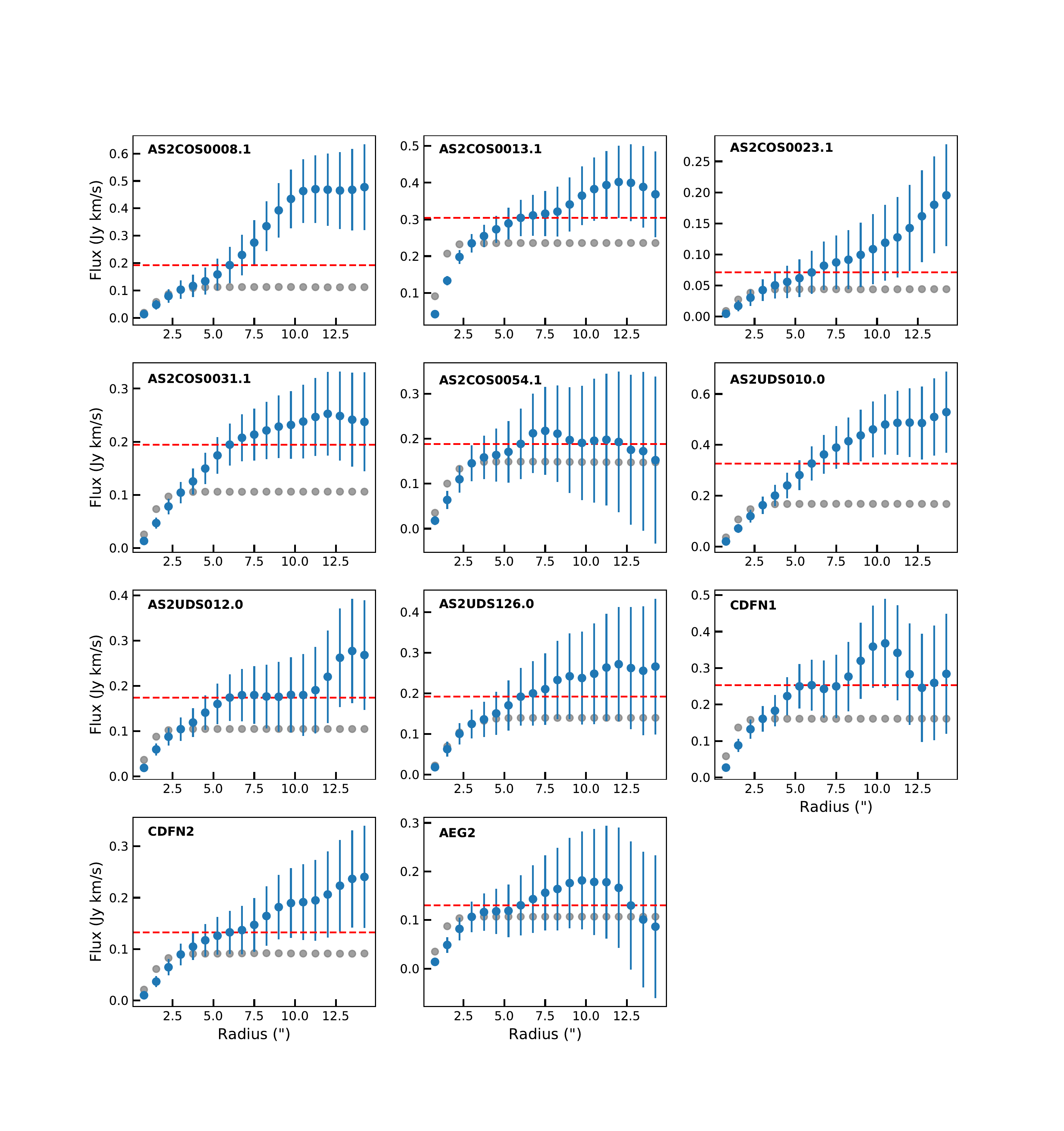}
    \caption{Curve of growth analysis performed on the detected sources, showing the flux recovered as a function of aperture radius. The dotted gray line is the curve of growth of the phase calibrator for each source. Compared to the calibrators, some of the brighter sources appear to be extended. The red dashed line marks the flux recovered within a 6$''$-radius aperture, where most of the sources converge. However, some of the fainter sources appear not to converge, which might be due to large-scale noise structures, more prominent due to the low SNR of the detections \citep{novak2020,AS2COSPEC2022}. Because of this, for the sources with integrated SNR$<$3, we extract the flux from an aperture 2.5$''$ in diameter and then correct to the total flux based on the median curve of growth. These are AS2COS0008.1, AS2COS0023.1, AS2COS0054.1,  AS2UDS126.0, AEG2 and CDFN8.}
    \label{cog}
\end{figure*}

We perform a curve-of-growth analysis on the 0th-moment maps to determine the optimal aperture to extract the line fluxes. We extract flux densities from a set of circular apertures of increasing diameter and determine the point at which the flux converges, shown in Figure \ref{cog}. Some of the fainter sources appear not to converge. This is due to the large-scale noise structures, more prominent due to the low SNR of the detections, which has also been observed in other data \citep{novak2020,AS2COSPEC2022}. We note that many of the brighter sources appear to be resolved compared to the phase calibrators, and show extended emission on roughly 6$''$-radius scales.

\subsection*{Non Detections}

\begin{figure}[!ht]
    \centering
    \includegraphics[scale=0.5]{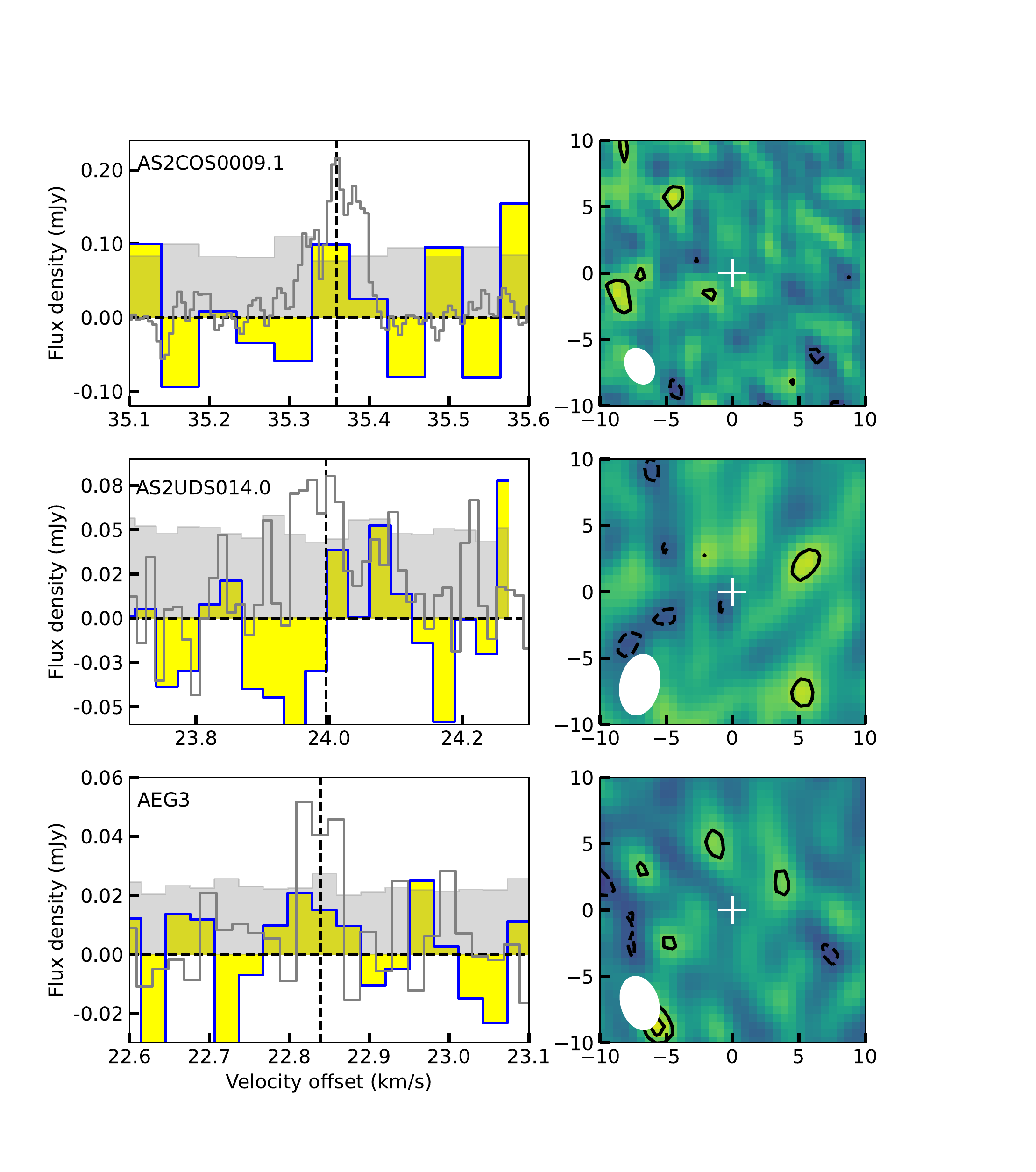}
    \caption{CO(1--0) spectra and 0th-moment maps of the three non-detections, AS2COS0009.1, AS2UDS014.0 and AEG3. The shaded region indicates the 1$\sigma$ rms per channel and the vertical dashed line indicates the expected redshift of the line based on the existing mid--$J$ CO line detections (shown by the gray spectra). To create the maps, we collapsed the data cubes over the line width of their corresponding mid--$J$ CO transition. Contours start at 2$\sigma$ and increase in steps of 1$\sigma$. The white cross marks the position of the peak emission of the mid--$J$ CO transition. The white ellipse indicates the FWHM of the beam.} 
    \label{non_detections}
\end{figure}

\bibliography{main}{}
\bibliographystyle{aasjournal}

\end{document}